\documentclass[a4paper,fleqn,usenatbib]{mnras}

\usepackage[T1]{fontenc}
\usepackage{ae,aecompl}

\usepackage{graphicx}	
\usepackage{amsmath}	
\usepackage{amssymb}	
\usepackage{subfigure}
\usepackage{xcolor}

\usepackage{mwe}
\newlength{\tempdima}
\newcommand{\rowname}[1]
{\rotatebox{90}{\makebox[\tempdima][c]{\textbf{#1}}}}

\title[Dust formation in embryonic pulsar-aided SNR]
{Dust formation in embryonic pulsar-aided supernova remnants}
\author[Omand et al.]
{Conor M. B. Omand$^{1}$\thanks{E-mail:omand@utap.phys.s.u-tokyo.ac.jp}, 
Kazumi Kashiyama$^{1,2}$, 
and Kohta Murase$^{3,4,5,6}$\\
$^{1}$Department of Physics, School of Science, the University of Tokyo, Tokyo 113-0033, Japan\\
$^{2}$Research Center for the Early Universe, the University of Tokyo, Tokyo 113-0033, Japan;\\
$^{3}$Department of Physics, The Pennsylvania State University, University Park, PA 16802, USA\\
$^{4}$Department of Astronomy \& Astrophysics, The Pennsylvania State University, University Park, PA 16802, USA\\
$^{5}$Center for Particle and Gravitational Astrophysics, The Pennsylvania State University, University Park, PA 16802, USA\\
$^{6}$Yukawa Institute for Theoretical Physics, Kyoto University, Kyoto 606-8502, Japan}

\date{Accepted XXX. Received YYY; in original form ZZZ}

\pubyear{2017}

\begin{document}
\label{firstpage}
\pagerange{\pageref{firstpage}--\pageref{lastpage}}
\maketitle

\begin{abstract}
We investigate effects of energetic pulsar wind nebulae (PWNe) on dust formation and evolution.  Dust emission has been observed in many supernova remnants that also have neutron stars as compact remnants. We study the dependence of dust formation time and size on properties of the ejecta and central pulsar. We find that a pulsar with an initial spin period $P \sim 1\mbox{-}10\,\rm ms$ and a dipole magnetic field $B \sim 10^{12\mbox{-}15}\,\rm G$ can either accelerate or delay dust formation, with a timescale of several months to over ten years, and reduce the average size of dust by a factor of $\sim$ 10 or more compared to the non-pulsar case.  We also find that infrared dust emission may be detectable in typical superluminous supernovae out to $\sim$ 100-1000 Mpc in 2-5 years after the explosion, although this depends sensitively on the spectral index of nonthermal emission from the nebula.  We discuss implications to previous supernova observations.  Some discrepancies between dust formation models and observations, such as the formation time in SN1987A or the dust size in the Crab Nebula, could be explained by the influence of a pulsar, and knowledge of the dust emission will be important for future ALMA observations of superluminous supernovae.

\end{abstract}


\section{Introduction}

In the expanding ejecta of a supernova, dust grains condense from cooling metal-rich gas.  These newly formed grains are injected into the interstellar medium (ISM), where they cause interstellar extinction and diffuse infrared emission, serve as a coolant and opacity source, catalyze H$_2$ formation, and serve as building blocks for planets and smaller rocky bodies.  

In particular, the origin of dust has been fiercely debated since the discoveries of a huge amount of dust grains at redshifts higher than $z$ = 5 \citep{2011A&ARv..19...43G}.  In the early universe, core-collapse SNe from massive stars are likely to be the dominant source of dust \citep{2007ApJ...662..927D}.  Infrared-submillimeter studies of SN1987A \citep{2011Sci...333.1258M, 2015ApJ...800...50M, 2014ApJ...782L...2I, 2012A&A...541L...1L, 2015ApJ...810...75D}, SNR G54.1+0.3 \citep{2017ApJ...836..129T}, Cas A \citep{2010ApJ...719.1553S, 2010A&A...518L.138B}, and the Crab Nebula \citep{2012ApJ...760...96G}, and several other supernova remnants (SNRs) \citep{2018arXiv181100034C}, as well as emission-line asymmetry studies of SN1980K, SN1993J, and Cas A \citep{2017MNRAS.465.4044B}, have reported a subsolar mass of cool dust formed in the ejecta which has not yet been destroyed by the supernova reverse shock \citep{2016A&A...590A..65M, 2016A&A...587A.157B}.  What fraction of the dust can survive the shock depends on their sizes after formation \citep[e.g.,][]{2007ApJ...666..955N, 2006ApJ...648..435N}, so understanding both the mass and size of dust produced in supernovae is important.

In previous dust formation studies, the effect of the pulsar wind nebula (PWN) emission has usually been neglected, even though nebulae have been found in several SNRs with dust.  In particular, the dust found in the Crab Nebula was smaller than predicted from models \citep{2009ASPC..414...43K, 2012ApJ...753...72T}; it is possible that this may have been due to the early PWN energy injection.  Dust formation in SN1987A also occurred later than predicted by most condensation models \citep{1993ApJS...88..477W, 1991A&A...249..474K, 2015A&A...575A..95S}, which may have been due to PWN emission, even though no compact object has yet been detected.  It is unknown if a PWN could be energetic enough to delay dust formation, yet be weak enough to remain below the detection limits.

Early supernova remnants have been searched for signals of a PWN, and recent optical studies of Type Ic SLSN 2015bn and peculiar Type Ib SN 2012au have presented tentative spectroscopic evidence of a central engine in the nebular phase \citep{2016ApJ...828L..18N, 2018ApJ...864L..36M}. Some studies have suggested X-rays \citep{kot+13,Metzger_et_al_2013, Kashiyama+16, MKM16} and gamma-rays  \citep{kot+13, MKM16} as possible probes for PWNe. 
X-ray studies have produced some tentative candidates \citep{Perna_Stella_2004,per+08,Margutti_et_al_17}, but are not largely constraining, and detecting gamma-ray signals should provide a more direct probe of the pulsar, but is more challenging and has not yet produced any candidates \citep{2015ApJ...805...82M,Renault-Tinacci:2017gon}.  
Recently, the idea has emerged to test the pulsar-driven model by detecting early radio and submillimetre PWNe emission \citep{MKM16} from broad-line Type Ic hypernovae and from Type Ic superluminous supernovae (SLSNe), which are both hypothesized by some to be pulsar-driven \citep[e.g.,][]{2010ApJ...724L..16P,qui+11,Inserra_et_al_2013,2014MNRAS.444.2096N,2016ApJ...828L..18N,Metzger_et_al_15,Wang_et_al_2015,Dai_et_al_2016}.  
Radio observations with facilities such as the Karl G. Jansky Very Large Array (VLA) are promising, but the ejecta attenuates signals at this wavelength for around 10-100 years \citep{MKM16,2018MNRAS.474..573O}, which roughly matches the age of our oldest SLSN candidates, so radio detections may still not be viable for a few years; recent studies have only placed weak constraints on the model \citep{2018MNRAS.473.1258S, 2018ApJ...857...72H}.  Submillimetre observations with facilities such as the Atacama Large Millimeter/submillimetre Array (ALMA) are also promising, as the ejecta only attenuates signals at this wavelength for around 1-10 years \citep{MKM16, 2018MNRAS.474..573O}. However, ALMA has previously been used to study dust in SNRs \citep[e.g.,][]{2014ApJ...782L...2I}, and dust emission in SLSN remnants may interfere with the detection of PWN emission.  Therefore, comparing dust emission spectra in SLSNe to PWN spectra would tell us if the dust will interfere with ALMA observations; be detectable in another band, such as infrared; or be subdominant to the PWN emission in all cases.

To study dust formation and emission in pulsar-driven supernovae, we use a steady-state model, which is overviewed in Section \ref{sec:thy}. So far, only the sublimation of previously formed dust has been studied \citep[e.g.,][]{2000ApJ...537..796W, 2011EP&S...63.1067K}.  The PWN emission can delay the formation of dust due to the added energy injection and is capable of sublimating dust as it forms, leading to longer formation times and the possible non-production of dust at all.  The PWN emission can also ionize the ejecta gas before dust formation, leading to increased temperature and Coulomb repulsion between ions, which may also prevent dust formation.  However, once dust has formed, the grains can absorb emission in the optical/UV band, greatly increasing their temperature compared to the case without a central pulsar.  These hot dust grains will re-emit in the infrared or submillimetre, and this emission might be detectable with telescopes like ALMA, Herschel, Spitzer, and the James Webb Space Telescope (JWST).  This gives an indirect signal, to compliment the direct radio and submillimetre detection discussed in \cite{MKM16} and \cite{2018MNRAS.474..573O}, by which we can detect newborn pulsars.

\section{Theory} \label{sec:thy}

\begin{figure}
\includegraphics[width=\linewidth]{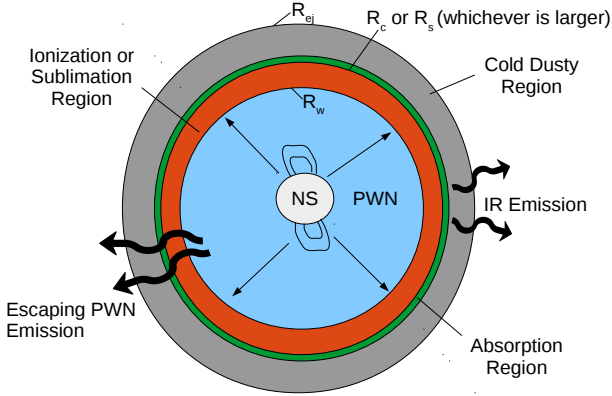}
\caption{The system examined in this paper; not to scale.  The PWN generated by the central NS pushes on and injects energy into the ejecta at $R_{\rm w}$, which we consider to have three layers: the sublimation or ionization region on the inside, where dust can not form due to non-thermal emission from the PWN; the absorption region with thickness $\ll$ $R_{\rm ej}$ at the edge of the sublimation or ionization region, where optical and UV photons are absorbed and infrared photons are emitted; and the cold/dusty region on the outside, where the dust is optically thin to infrared emission and is not heated by non-thermal emission.  Note that all three regions do not always appear, depending on parameters and time evolution.}
\label{fig:cartoon}
\end{figure}

The system we consider is shown schematically in Figure~\ref{fig:cartoon}, and we examine it from the supernova explosion until the beginning of the Sedov phase, when a reverse shock is driven back into the ejecta and is expected to destroy smaller dust grains via sputtering.  The spin down of the neutron star generates a PWN which pushes against and injects energy into the ejecta at $R_{\rm w}$, the edge of the shocked wind region.  We use the one zone model approximation, where all the ejecta is contained between $R_{\text{w}}$ and $R_{\text{ej}}$.  The inner region of the ejecta can be either a sublimation region of radius $R_c$, if the optical/UV luminosity is high enough to heat the dust above its sublimation temperature, or an ionization region of radius $R_s$, if high energy radiation can ionize most of the gas in the region.  Both of these possibilities should prevent dust formation in that region; we discuss the conditions for these regions in Sections \ref{sec:dustsub} and \ref{sec:gasio}.  Outside this region, there will be a thin region where most of the optical/UV emission will be absorbed by dust and re-emitted in the infrared; this absorption region has a thickness given by $\tau_{\rm opt/UV} \sim$ 1 and it's emission is described in Section \ref{sec:dustem}.  Outside of the absorption region is the cold, dusty region, where the dust is not being heated by non-thermal radiation, and cools via adiabatic and radiative cooling; we neglect the emission of this region entirely, since it is expected to be much cooler than the absorption region.  If $R_c$ or $R_s$ is $>$ $R_{\text{ej}}$, then no dust will form and the entire ejecta will be the sublimation/ionization region with no absorption or cold, dusty region.

\subsection{Pulsar Spin Down, Ejecta Dynamics, and Non-Thermal Emission}

To calculate pulsar spin-down, ejecta dynamics, and non-thermal emission we use the model from \cite{Kashiyama+16}.

The velocity of the ejecta $v_{\rm ej}$ is calculated from the kinetic energy $E_K = M_{\rm ej}v_{\rm ej}^2/2$, which evolves with time as
\begin{equation}
\frac{dE_K}{dt} = \frac{E_{\rm int}}{t_{\rm dyn}}.
\label{eqn:dekrej}
\end{equation}
Here $M_{\rm ej}$ is the mass of the ejecta, $E_{\rm int}$ is the total internal energy, and $t_{\rm dyn} = R_{\rm ej}/v_{\rm ej}$ is the dynamical timescale of the ejecta, where $R_{\rm ej}$ is the outer radius of the ejecta.  
Note that $E_{\rm int}$ is determined by the balance between heating via the PWN irradiation and radioactive decay of $^{56}$Ni and $^{56}$Co and cooling via the radiative energy loss and expansion.
The radius of the inner edge of the ejecta $R_{\rm w}$ increases as:

\begin{equation}
v_{\rm w} = \frac{dR_{\rm w}}{dt} = v_{\rm nb} + \frac{R_{\rm w}}{t}
\label{eqn:drwdt}
\end{equation}
where $v_{\rm nb}$, the velocity of the shocked wind region, is determined by the pressure balance at $R_{\rm w}$ via

\begin{equation}
V_\text{nb} \approx \sqrt{\frac{7}{18}\frac{\int L_\text{SD} \times \min[1,\tau^\text{nb}_\text{T}v_\text{nb}/c]dt}{M_\text{ej}}\left(\frac{R_\text{ej}}{R_\text{w}}\right)^3},
\label{eqn:vnb}
\end{equation}
where the factor $\min[1,\tau^\text{nb}_\text{T}v_\text{nb}/c]$ is the fraction of spin-down luminosity deposited in the SN ejecta and $\tau^\text{nb}_\text{T}=(R_\text{w}/R_\text{ej})\tau_\text{ej}$, where $\tau_\text{ej}$ is the optical depth of the ejecta,

\begin{equation}
\tau_\text{ej} = \frac{3\kappa M_\text{ej}}{4\pi R^2_\text{ej}},
\label{eqn:tauejesc}
\end{equation}
where $\kappa$ is the Thompson opacity.

We assume that the ejecta mass $M_{\rm ej}$ is confined in a volume of $V_{\rm ej} = (4\pi/3)(R_{\rm ej}^3-R_{\rm w}^3)$ with a uniform density $\rho_{\rm ej} = M_{\rm ej}/V_{\rm ej}$; the density profile is always constant, regardless of the compression of the ejecta due to the PWN.  The initial conditions of the model have $R_{\rm ej} = 1.0 \times 10^{11}$ cm and $R_{\rm w} = 0.1R_{\rm ej}$, although our results are not sensitive to these values.
In particular, for cases with rapidly rotating pulsars, the expanding nebula compresses the ejecta into a thin shell ($R_{\rm ej}-R_{\rm w} \ll R_{\rm ej}$).  Here, if $R_{\rm w} \geq 0.8R_{\rm ej}$, we set $R_{\rm w} = 0.8R_{\rm ej}$; this value gives a shock compression ratio of 5, which is in between the values of 4 and 7 for adiabatic and isothermal/radiative shocks respectively.  For simplicity, we take this value due to the uncertainty of the nature of the shock. Also note that once a pulsar injects energy comparable to the initial explosion energy into the ejecta, the ejecta will undergo a hot bubble breakout~\citep{2017MNRAS.466.2633S}. This can modify the density structure especially of the outer ejecta, which is not taken into account in our model. 

In general, embryonic PWN spectra are obtained by solving kinetic equations that take into account electromagnetic cascades in the nebula~\citep{2015ApJ...805...82M}. 
For simplicity, in this work, the fiducial PWN spectrum is approximated to be:

\begin{equation}
\nu F_\nu = \frac{\epsilon_{\rm e} L_{\rm SD}(t)}{{\cal R}_b}
\begin{cases}
(E_\gamma/E^b_\text{syn})^{2-\alpha_1} & (E_\gamma < E^b_\text{syn}), \\
(E_\gamma/E^b_\text{syn})^{2-\alpha_2} & (E^b_\text{syn} < E_\gamma),
\end{cases}
\label{eqn:wns}
\end{equation}
where $\alpha_1 = 1.5-1.8$ and $\alpha_2 = 2.15$ unless otherwise noted,  and $\epsilon_{\rm e}$ is the fraction of spin-down energy that goes into the emission, which we take to be a free parameter of order unity.\footnote{In the very early phase with a high compactness of the source, Compton cascades are fully developed in the so-called saturated regime, in which a flat energy spectrum is expected \citep[e.g.][]{Metzger_et_al_2013}. At later phases, the spectrum has two humps of synchrotron and inverse-Compton emissions \citep{2015ApJ...805...82M}.  Because we focus on the dust-forming phase, which happens at longer timescales, we should take the above values expected at low frequencies.}
In the fast cooling limit, one has $\alpha_{1,2}=(2+q_{1,2})/2$ for the electron-positron injection index $q_{1.2}$. 
Also, $L_{\rm SD}$ is the total electromagnetic emission from the spin-down of the pulsar

\begin{equation}
L_{\rm SD}(t) \approx L_{\rm SD,0}f_{\rm SD}(t),
\label{eqn:lsd}
\end{equation}
where

\begin{align}
L_{\rm SD,0} &\simeq 8.6 \times 10^{45} \text{ erg/s } \left(\frac{B}{10^{13}\text{ G}}\right)^{2} \left(\frac{P}{1 \text{ ms}}\right)^{-4}, \\
f_{\rm SD}(t) &= \begin{cases}
1 & (t < t_{\rm SD}), \\
(t/t_{\rm SD})^{-2} & (t > t_{\rm SD}),
\end{cases} \\
t_{\rm SD} &\simeq 37 \text{ days } \left(\frac{B}{10^{13}\text{ G}}\right)^{-2} \left(\frac{P}{1 \text{ ms}}\right)^{2},
\end{align}
where $B$ and $P$ are the initial rotation period and dipole magnetic field of the pulsar, which we vary; ${\cal R}_b$ is a normalization factor given by
${\cal R}_b \sim1/(2-\alpha_1)+1/(\alpha_2-2)$ and the break photon energy is

\begin{equation}
E^b_\text{syn} = \frac{3}{2} \hbar \gamma^2_b \frac{eB_{\rm PWN}}{m_{\text{e}}c},
\label{eqn:ebsyn}
\end{equation}
where $\gamma_b = 3 \times 10^5$ (we discuss the uncertainty of this parameter in Section \ref{sec:disgammab}) and $B_{\rm PWN}$ is the magnetic field in the PWN

\begin{equation}
B_{\rm PWN}^2 \approx \frac{6\epsilon_B L_{\rm SD,0}}{v_{\rm w}^3 t_{\rm SD}^2}
\begin{cases}
(t/t_{\rm SD})^{-2} & (t < t_{\rm SD}), \\
(t/t_{\rm SD})^{-3} & (t > t_{\rm SD}),
\end{cases}
\label{eqn:bpwn}
\end{equation}
where $\epsilon_B = 3 \times 10^{-3}$ is the fraction of spin-down energy that goes into the PWN magnetic energy.  
At $t = t_{\rm SD}$, 
this has a value of:

\begin{equation}
B_{\rm PWN} (t_{\rm SD})\sim 113 \text{ G } \left(\frac{v_{\rm w}}{10^9 \text{ cm s$^{-1}$}}\right)^{-3/2} \left(\frac{B}{10^{13}\text{ G}}\right)^{3} \left(\frac{P}{1 \text{ ms}}\right)^{-4}.
\label{eqn:bpwntsd}
\end{equation}
This field evolution assumes that the field energy $B_{\rm PWN}^2$ is proportional to the total energy of the PWN. 
Although it may not be completely true, such an assumption can be justified when the magnetic field is toroidally dominated.

This assumed spectrum is motivated by previous studies~\citep{2015ApJ...805...82M, MKM16,2018MNRAS.474..573O}; which assumes a broken power-law injection of relativistic electrons and positrons in to the PWN, inferred from Galactic PWN observations~\citep{tt10, tt13}, and calculates synchrotron emission and inverse Compton scattering, pair production and consequent electromagnetic cascades. 
Equation \ref{eqn:wns} can be a good approximation especially for the early phase where all the injected electrons and positrons are in the fast cooling regime. However, this is not necessary the case in the late phase; relatively low-energy electrons and positrons are in the slow cooling regime. As a result, in general, the fast-cooling limit spectrum overestimates the PWN flux at low-energy bands, e.g., in the infrared, submm, and radio bands. In more realistic cases, previously injected electrons (sometimes called relic electrons) can change the power-law spectrum~\citep{MKM16}. We discuss effects of these relic electrons and uncertainties in the PWN spectrum in Appendix \ref{sec:disrealspec}. 
The non-thermal flux can be reduced if a significant fraction of the spin-down energy remains unconverted to radiation or $\epsilon_e$ is decreased, but the optical flux should also be reduced correspondingly. 
Note that only the luminosity in the optical band is important for dust temperature and sublimation.

Throughout this paper, we take the ejected nickel mass $M_{\text{Ni}}$, 
the initial kinetic energy of the ejecta $E_{\text{SN}}$, and the opacity $\kappa$ to be 0.1 $M_{\sun}$, $10^{51}$ erg, and 0.1 g cm$^{-2}$ respectively, as in \cite{2018MNRAS.474..573O}.  

\subsection{Dust Formation}

Dust formation in SN ejecta has mainly been studied with classical nucleation theory and its extension \citep{1989ApJ...344..325K, 1991A&A...249..474K, 2003ApJ...598..785N, 2008ApJ...684.1343N, 2010ApJ...713..356N, 2011ApJ...736...45N, 2001MNRAS.325..726T, 2007MNRAS.378..973B}.  In this theory, dust condensation is described by the formation of stable seed nuclei and their growth, where the formation rate is derived by assuming the nucleation current to be in a steady state \citep{2013ApJ...776...24N}.  This theory has allowed us to predict the size distribution and mass of condensing grain species, and these results have nicely explained the mass of dust formed in SN1987A \citep{1991A&A...249..474K} and the formation and evolution processes of dust in Cas A \citep{2010ApJ...713..356N}.

The model we use for dust formation is the steady-state model, first developed by \cite{1987PThPh..77.1402K} by introducing the concept of a key species or key molecule, which has the lowest collisional frequency among gaseous reactants, and then generalized by \cite{2013ApJ...776...24N}, whose formulation we take here.  In this formulation, collisions between gaseous key molecules and clusters of $n$ key molecules, which we refer to as $n$-mers, control the reaction kinetics.

As the gas cools, dust condensation proceeds via the formation of clusters and subsequent attachment of key molecules to those clusters.  The concentration of gas $c_1$ (we denote the concentration of $n$-mers $c(n,t) = c_n$) is given by
\begin{equation}
c_1 = \frac{M_{\rm ej}f_{\text{KM}}(1 - f_{\text{con}})}{V_{\rm ej} m_1}
\label{eqn:cevo}
\end{equation}
\noindent
where 
$f_{\text{KM}}$ is the initial mass fraction of the key molecule in the ejecta, $f_{\text{con}}$ is the condensation efficiency, and $m_1$ is the mass of the key molecule.  

The growth rate of grains, which we assume are spherical, is given by
\begin{equation}
\frac{da}{dt} = s\Omega_0 \left( \frac{kT_{\text{gas}}}{2\pi m_1}\right)^{\frac{1}{2}}c_1\left( 1 - \frac{1}{S}\right),
\label{eqn:cevo}
\end{equation}
\noindent
where $a$ is the grain radius, $s$ is the sticking probability of the key molecule onto grains, $\Omega_0$ is the volume of the condensate per key molecule, $k$ is the Boltzmann constant, $T_{\text{gas}}$ is the gas temperature, and $S$ is the supersaturation ratio
\begin{align}
\ln S = \frac{A}{T_{\text{gas}}} - B + \ln \left( \frac{c_1kT_{\text{gas}}}{p_s}\right) + \ln \Xi,
\label{eqn:ssr}
\end{align}
\noindent
where $A$ and $B$ are thermodynamic constants given in \cite{2003ApJ...598..785N}, 
$p_s = 1\,{\rm bar} = 10^6\,{\rm erg\,cm^{-3}}$, and  
\begin{equation}
\Xi = \frac{\prod_{k=1}^i (p_k^{\mathcal A}/p_s)^{\nu_k}}{\prod_{k=1}^j (p_k^{\mathcal B}/p_s)^{\eta_k}},
\label{eqn:xidef}
\end{equation}

\noindent
where $\nu_k$ and $\eta_k$ are the stoichiometric coefficients and $p_k^{\mathcal A}$ and $p_k^{\mathcal B}$ ($k = 1$-$i$ and $1$-$j$ respectively) are the partial pressures for the gaseous reactants and products, $\mathcal{A}_k$ and $\mathcal{B}_k$, respectively, in the general chemical reaction below
\begin{equation}
\mathcal{Z}_{n-1} + ( \mathcal{X} + \nu_1\mathcal{A}_1 + ... + \nu_i\mathcal{A}_i ) \rightleftharpoons \mathcal{Z}_n + (\eta_1\mathcal{B}_1 + ... + \eta_j\mathcal{B}_j),
\label{eqn:chemreac}
\end{equation}

\noindent
where $\mathcal{Z}_n$ is an $n$-mer cluster generated from the nucleation of $n$ key molecules $\mathcal{X}$.

In the steady-state approximation, the current density $J_n$ is independent of $n$, being equal to the steady-state nucleation rate $J_s$.  Starting from the reaction equation above, and following the derivation from \cite{2013ApJ...776...24N}, the steady-state nucleation rate is
\begin{equation}
J_s=s\Omega_0 \left( \frac{2\sigma_{\text{ten}}}{\pi m_1}\right)^{\frac{1}{2}}c_1^2 \Pi \exp \left( -\frac{4}{27} \frac{\mu^3}{(\ln S)^2} \right),
\label{eqn:jsdef}
\end{equation}
\noindent
where $\sigma_{\text{ten}}$ is the surface tension of the condensate taken from \cite{2003ApJ...598..785N}, $\mu = 4\pi a_0^2 \sigma_{\text{ten}}/kT_{\text{gas}}$ is the ratio between the surface energy of the condensate due to tension and the thermal energy of the gas, $a_0 = (3 \Omega_0/4 \pi)^{1/3}$ is the hypothetical grain radius per key molecule, which is calculated in \cite{2003ApJ...598..785N}, and the correction factor $\Pi$ is given by 
\begin{equation}
\Pi = \left( \frac{\prod_{k=1}^i (c_k^{\mathcal{A}}/c_1)^{\nu_k}}{\prod_{k=1}^j (c_k^{\mathcal{B}}/c_1)^{\eta_k}} \right)^{\frac{1}{\omega}},
\label{eqn:pidef}
\end{equation}
\noindent
where 
\begin{equation}
\omega = 1 + \sum_{k=1}^i \nu_k - \sum_{k=1}^j \eta_k.
\label{eqn:omedef}
\end{equation}

Once $J_s$ is calculated, dividing by $\tilde{c}_1$ gives us $I_s$, which is used to calculate
\[
\frac{dK_i}{dt} =
\begin{cases}
I_s(t)n_*^{\frac{i}{3}} + \frac{i}{a_0} \left( \frac{da}{dt} \right) K_{i-1} & \text{for $i=1-3$} \\
I_s(t) & \text{for $i=0$}.
\end{cases}
\label{eqn:dkdt}
\]
Here $K_0$ represents the number density of dust grains ($K_0 = n_{\text{dust}}/\tilde{c}_1$), and $K_3$ represents the number fraction of key molecules locked in dust grains.  Therefore, we can calculate the condensation efficiency $f_{\text{con}}(t)$ and average radius $a_{\text{ave}}(t)$ by
\begin{align}
f_{\text{con}} =& K_3, \\
a_{\text{ave}} =& a_0 \left( \frac{K_3}{K_0}\right)^{\frac{1}{3}}.
\label{eqn:fconaave}
\end{align}

We tested the formation of dust for two initial dust compositions (mass fractions are given in Table~\ref{tbl:comps}), which we call the Ib and Ic compositions due to the supernovae they correspond to.  These compositions are based on nucleosynthesis calculations \citep{2010ApJ...725..940Y} used in recent radiative transfer simulations of various types of supernovae with various types of progenitors, which account for nuclear fusion during the explosion \citep{2011MNRAS.414.2985D, 2012MNRAS.426L..76D, 2015MNRAS.453.2189D, 2016MNRAS.458.1253V, 2017A&A...603A..51D}.  The Ib composition is similar to that of a small (ZAMS mass of 15-25 M$_{\odot}$) Wolf-Rayet star in a binary with roughly solar metallicity; one would expect about 3-5 M$_{\odot}$ of ejecta in this case. The Ib composition is also fairly similar to a low metallicity Wolf-Rayet star without a binary companion with a ZAMS mass of around 25 M$_{\odot}$; the ejecta mass in this case would be $\sim$ 15 M$_{\odot}$.  The Ic composition is similar to that of a large solar metallicity Wolf-Rayet star with ZAMS mass of around 60 M$_{\odot}$ evolved without a binary companion; one would expect about 5-7 M$_{\odot}$ of ejecta in this case.

The biggest differences between the two are the lack of Si in the Ic composition and the lower overall numbers in the Ib composition.  While the Si mass fraction is not zero in real SNe, the simulations give a mass fraction of about 10 times lower than that of Mg for the Ic composition progenitor; this is small enough where we expect MgO grains to be formed in much greater quantity than MgSiO$_3$ or Mg$_2$SiO$_4$, so we neglect Si completely for the Ic composition.  The Ib composition has lower numbers because a large fraction of the gas is still He, which does not form dust and is thus neglected in this study.  The large fraction of He means that observed SNe with the Ib composition would be either Type Ib or IIb, depending on if any H gas still remained as well, while observed SNe with the Ic composition would be seen as Type Ic.

We examine two different types of dust growth for each composition.  For the Ib composition we examine the formation of C and MgSiO$_3$ grains, which we expect to be formed preferentially over Mg$_2$SiO$_4$ by about a factor of 3 \citep{2010ApJ...713..356N}.  For the Ic composition, since there is not enough Si to form large quantities of MgSiO$_3$ or Mg$_2$SiO$_4$, we examine growth of C and MgO grains.  The growth reaction equations for these clusters are
\begin{align}
\text{C}_{n-1} + \text{C} & \rightleftharpoons \text{C}_n, \label{eqn:creac} \\
\text{MgSiO}_{3,n-1} + \text{Mg} + \text{SiO} + \text{O} & \rightleftharpoons \text{MgSiO}_{3,n}, \label{eqn:mgsio3reac} \\
\text{MgO}_{n-1} + \text{Mg} + \text{O} & \rightleftharpoons \text{MgO}_n . \label{eqn:mgoreac}
\end{align}
The physical properties of each dust grain used in the calculation are listed in Table~\ref{tbl:gprop}.  We assume for the Ib composition that the concentrations of Mg and Si gas remain equal, and we assume that the number of oxygen atoms remains fixed, since the ejecta is oxygen dominated and grain formation will not significantly affect the concentration.  

\begin{table}
\begin{tabular}{|c|c|c|c|c|} \hline
Composition & $f_{\text{C}} $ & $f_{\text{O}}$ & $f_{\text{Mg}}$ & $f_{\text{Si}}$ \\ \hline 
Ib & 0.1 & 0.3 & 0.03 & 0.03 \\
Ic & 0.3 & 0.6 & 0.05 & 0 \\ \hline 
\end{tabular}
\caption{Initial mass fractions of the different gaseous elements in the ejecta.}
\label{tbl:comps}
\end{table}

\begin{table}
\begin{tabular}{|c|c|c|c|} \hline
Grain Type & C$_{\text{(s)}}$ & MgSiO$_{3\text{(s)}}$ &  MgO$_{\text{(s)}}$ \\
Key Species & C$_{\text{(g)}}$ & Mg$_{\text{(g)}}$ & Mg$_{\text{(g)}}$  \\
$A/10^4$ (K) & 8.64726 & 25.0129 & 11.9237  \\
$B$ & 19.0422 & 72.0015 & 33.1593  \\
$a_0$ ({\AA}) & 1.281 & 2.319 & 1.646 \\
$\sigma_\text{ten}$ (erg cm$^{-2}$) & 1400 & 400 & 1100 \\ \hline 
\end{tabular}
\caption{The properties of the dust grains considered in this study.  The subscript (s) and (g) represent solids and gasses respectively.  Since Mg and Si have the same concentration in the Ib composition, either one can be used as the key species.  Values are taken from \protect\cite{2003ApJ...598..785N}.}
\label{tbl:gprop}
\end{table}

We ignore the formation of CO molecules, even though in oxygen-rich ejecta (which both compositions have) it is expected that carbon dust will not form in large quantities due to the preferential formation of CO molecules.  Since our model is a one-zone model, including CO formation would mean that carbon dust formation would be greatly suppressed.  In more complicated models, supernovae have both an oxygen-rich shell where silicate and Mg-molecule-based dust formation is dominant and a carbon-rich shell where carbon dust formation is dominant \citep[e.g.,][]{2008ApJ...684.1343N,2010ApJ...713..356N}.  For most supernovae, we would only expect carbon dust formation in the carbon-dominant shell, which surrounds the oxygen-rich shell and usually contains $\sim$ 50\% of the carbon atoms, but particularly in SLSNe, turbulent mixing mixes the gas and homogenizes the ejecta, meaning that carbon dust will not form.  For this reason, we treat the formation of each species independently, without accounting for shielding due to the early formation of one type of dust.

We take $s = 0.8$ and $n_* = 100$, the sticking probability of a colliding gas molecule and the minimum number of molecules for a cluster to be considered a dust grain respectively, for all calculations; as long as $n_*$ is large enough, the results do not qualitatively change - this is discussed in Appendix B of \cite{2013ApJ...776...24N}.

\subsection{Dust Sublimation}\label{sec:dustsub}

Once the gas has cooled enough to form, the dust can still be sublimated by the PWNe optical-UV emission.  The equation for dust grains in radiative  equilibrium between absorbing PWN emission and emitting thermal emission in the IR band is
\begin{equation}
\frac{L_{\text{opt/UV}}}{4\pi r^2}Q_{\text{opt/UV}}\pi a^2 = \langle Q \rangle _T 4\pi a^2 \sigma T_{\text{dust}}^4, 
\label{eqn:graineq}
\end{equation}

\noindent
where $L_{\text{opt/UV}}$ is the non-thermal luminosity in the band between 2-6 eV (0.2-0.6 $\mu$m), $\sigma$ is the Stefan-Boltzmann constant, $r$ is the radius of the dust grain's position, $Q_{\text{opt/UV}}$ is the absorption efficiency factor averaged over the optical/UV spectrum, which we assume is $\approx$ 1, and finally
\begin{align}
\langle Q \rangle _T = & \frac{\int B_{\nu}(T_{\text{dust}})Q_{\text{abs,}\nu}d\nu}{\int B_{\nu}(T_{\text{dust}})d\nu} \\
\approx & \frac{Da_{-5}(T_{\text{dust}}/2300 \text{ K})}{1+Da_{-5}(T_{\text{dust}}/2300 \text{ K})}, 
\label{eqn:qt}
\end{align}

\noindent
where $a_{-5} = a/10^{-5}$ cm and $D$ is a constant ($\sim$ 0.3 for C dust grains, $\sim$ 0.03 for silicates and MgO) \citep{1984ApJ...285...89D}.  These choices of emissivities are consistent with studies examining the sublimation of previously formed dust grains larger than $10^{-5}$ cm \citep{2000ApJ...537..796W}, but is not completely accurate as the dust is growing, or if the dust does not grow to $10^{-5}$ cm.  \cite{1984ApJ...285...89D} calculated the emissivities of both graphite and silicates using their dielectric function and found it varied strongly and non-linearly with both grain size and absorption frequency.  We discuss this approximation further in Section \ref{sec:dis}.

Dust will be sublimated if its equilibrium temperature is greater than the critical temperature $T_c$ for supersaturation, which can be calculated by setting $S=1$ in Equation~\ref{eqn:ssr}.  From Equation~\ref{eqn:graineq}, the critical radius for dust sublimation is:
\begin{equation}
R_c = \left(  \frac{L_{\text{opt/UV}}}{16\pi \sigma T_c^4} \frac{Q_{\text{opt/UV}}}{\langle Q \rangle _{T_c}}\right)^{\frac{1}{2}}.
\label{eqn:tdust}
\end{equation}

If $R_c < R_{\text{ej}}$ (the edge of the ejecta), then no dust can be formed due to sublimation from the PWNe emission.  Any dust that would have formed at this point is converted back to gas in our calculation.

\subsection{Dust Emission}\label{sec:dustem}

Once dust can start to form without being sublimated, it emits thermally in the infrared band.  The optical-UV optical depth is:
\begin{equation}
\tau_{\text{opt/UV}} = \int_{R_c}^{R_{\text{ej}}} n_{\text{dust}}(r) \pi a^2 dr
\label{eqn:uvopdep}
\end{equation}

\noindent
is $\gg$ 1 in the dusty ejecta, so only a thin layer (the absorption region) will be heated by the PWN emission.  This region will be located just outside $R_c$ and will emit just below $T_c$ if $R_c > R_{\text{w}}$, or be located at $R_{\text{w}}$ and emit at $T_\text{dust}(R_{\text{w}})$, with a blackbody luminosity, 
\begin{equation}
L_{\nu} = 4\pi R^2 \langle Q \rangle _T \pi \frac{2h\nu^3}{c^2} \frac{1}{e^{\frac{h\nu}{k_BT}}-1}.
\label{eqn:thicklv}
\end{equation}
Although the reprocessed emission is sometimes modelled with a frequency-dependent emissivity with $L_{\nu} \propto \nu^{2+\beta}$ for $h\nu < kT$ \citep[e.g.,][]{1991ApJ...381..250B, 2007ApJ...663..866D, 2010ApJ...708..127S}, we use the frequency-averaged emissivity from Equation~\ref{eqn:qt}.  Since we are only interested in the peak of the spectrum, the exact spectral index in the Rayleigh-Jeans limit is not important to our results.

Since the reprocessed emission lies at longer wavelengths than the absorbed PWN emission, which are longer than the typical size of the dust grains, the rest of the dust will appear optically thin for this dust emission. The thermal emission at $T_c$ or $T_\text{dust}(R_{\text{w}})$ will be directly observable.

\subsection{Gas Ionization}\label{sec:gasio}

Ionization of the gas can cause a temperature increase due to the collisions with free electrons, as well as increased Coulomb repulsion between charged ions which may prevent dust formation.  However, ion-molecule reactions proceed more quickly due to ions inducing dipole moments in neutral atoms, enhancing their electrostatic attraction \citep{2005fost.book.....S}.  Although the extent to which these effects compete and the ionization states in which they dominate are not well known, it is important to identify the region in which they will be important.

We calculate the radius out to which the gas can be ionized by the non-thermal radiation using the standard formula for the Str{\"o}mgen Radius $R_s$, but slightly modified due to the ejecta being in a shell from $R_{\text{w}}$ to $R_{\text{ej}}$.  The formula becomes 
\begin{equation}
R_s = \left(\frac{3}{4\pi}\frac{\Phi}{c_1^2 \beta_2} + R_{\text{w}}^3 \right)^{\frac{1}{3}},
\label{eqn:stromrad}
\end{equation}
where $\beta_2$ is the total recombination rate, which depends on electron temperature and chemical composition, and $\Phi$ is the flux of ionizing photons from the source.  $\Phi$ can be calculated from the spectrum in Equation~\ref{eqn:wns}, by
\begin{align}
& \Phi= \frac{F^b_{\nu}}{E^b_\text{syn}} \times \nonumber\\
& \begin{cases} 
\frac{1}{\alpha_1-1}[(\frac{E_I}{E^b_\text{syn}})^{-(\alpha_1-1)} -1] + \frac{1}{\alpha_2-1} & (E_I < E^b_{\rm syn}), \\
\frac{1}{\alpha_2-1}\left(\frac{E_I}{E^b_\text{syn}}\right)^{-(\alpha_2-1)} & (E^b_\text{syn} < E_I)
\end{cases}
\label{eqn:ionflux}
\end{align}
where $E_I$ is the ionization energy of the gas atom, which depends on the atoms being ionized, but is between 5-15 eV for all atoms of interest here.  

For gas density we use the concentration if no dust is formed $\tilde{c}_1$, since $R_s$ is not physically relevant to this study after dust is formed, but we examine multiple types of dust and want to treat their formation and ionization independently.  We do not couple this calculation to the dust formation and sublimation calculation, as it is not well known what effect partial ionization of the ejecta will have on dust formation.

This formulation produces results that are mostly consistent with recent results by \cite{2018arXiv180605690M}, who calculated the ionization state of hydrogen- and oxygen-rich ejecta in a system with a $B$ = 10$^{14}$ G and $P$ = 1 ms rotating pulsar and 10 M$_{\sun}$ of ejecta.  They find that the density averaged ionization fraction increases slowly for the hydrogen-rich ejecta and stays roughly constant for oxygen-rich ejecta. 
However, their ejecta profile consists of a homogeneous core below $R_{\text{w}}$ and a high-velocity tail instead of a thin shell from $R_{\text{w}}$-$R_{\text{ej}}$, like ours, which will change the fraction of ejecta which becomes ionized.
Note that the gas ionization is important for the detectability of radio and submm emission because of various absorption processes.  \cite{MKM16} studied effects of synchrotron self-absorption and free-free absorption, and found that the free-free absorption becomes irrelevant for $\gtrsim3-30$~yr ($\gtrsim1-3$~yr) at GHz (at 100~GHz) assuming the singly ionized state~\citep{MKM16}, which has been confirmed by \cite{2018arXiv180605690M}.

\section{Results}

We perform a parameter study for the initial pulsar rotation period $P$ and the initial magnetic field $B$.  The overall PWN flux is multiplied by the factor $\epsilon_e$, which is analogous to changing the power law spectrum or the dust absorption bandwidth, since only the total luminosity in the optical band is important for dust temperature and sublimation (Sec. \ref{sec:dustem}).  We investigate five sets of ejecta and PWNe parameters, shown in Table~\ref{tbl:planrun}; they will give us qualitative information on the effect of changing ejecta mass, the PWNe spectrum, and the composition as well as being case studies for typical binary Wolf-Rayet progenitors (Ib5-1) and low metallicity single progenitors (Ib15-1), while Ic5-1 will be a case study for large solar metallicity single progenitors and Ic15-1 will be a case study for millisecond pulsar-driven superluminous supernovae.  It's also worth noting that Ib15-1 and Ic5-1 have the same amount of total carbon, so comparing these two give insight into the effects of significant changes to the dynamics.

\begin{table}
\begin{tabular}{|c|c|c|c|} \hline
ID & Composition & $M_{\rm ej}$ M$_{\sun}$ & $\epsilon_e$ \\ \hline 
Ib5-1 & Ib & 5 & 1 \\
Ib5-05 & Ib & 5 & 0.5 \\
Ib15-1 & Ib & 15 & 1 \\
Ic5-1 & Ic & 5 & 1 \\ 
Ic15-1 & Ic & 15 & 1 \\ \hline 
\end{tabular}
\caption{The five sets of ejecta and PWNe parameters we study.  $\epsilon_e$ is a multiplying factor for the PWNe flux.}
\label{tbl:planrun}
\end{table}

\subsection{Effects of a Pulsar}

\begin{figure}
\begin{subfigure}
  \centering
  \includegraphics[width=\linewidth]{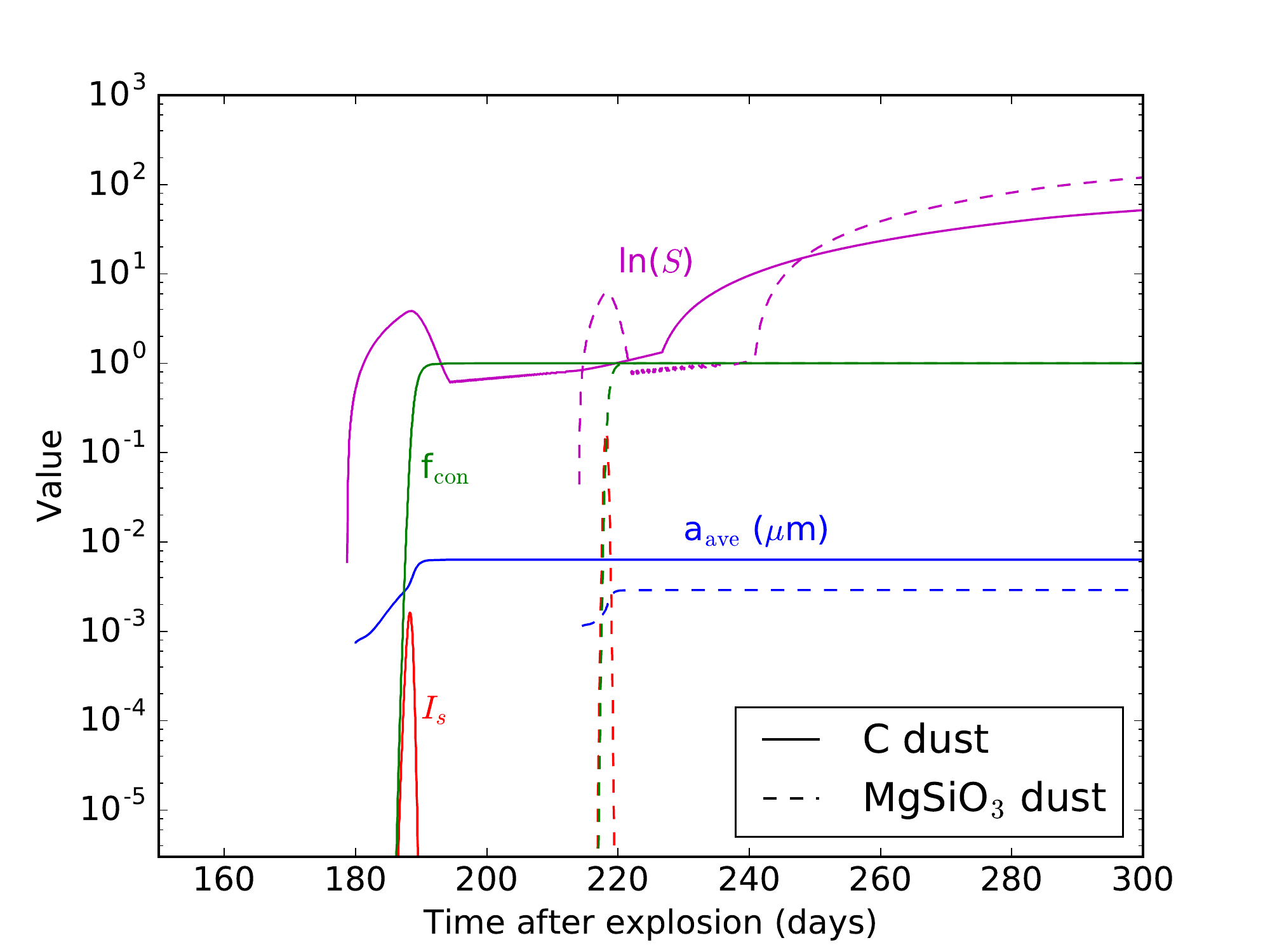}
\end{subfigure} \\
\begin{subfigure}
  \centering
  \includegraphics[width=\linewidth]{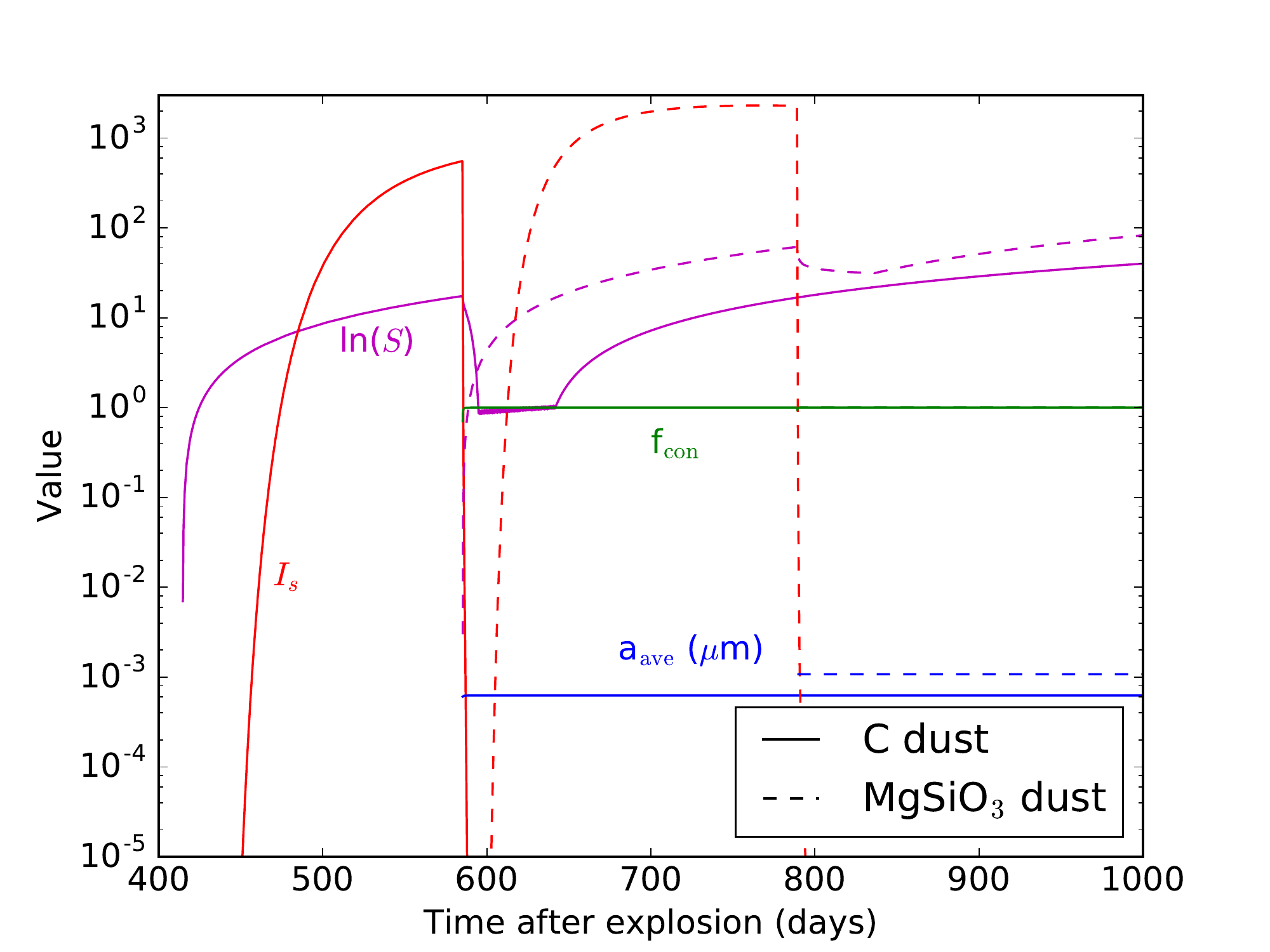}
\end{subfigure} \\
\caption{The time evolution of $\ln(S)$, $I_s$, $f_{\text{con}}$ and $a_{\text{ave}}$ for both C and MgSiO$_3$ dust in the Ib5-1 composition without a pulsar (top) and with a $P$ = 2 ms, $B$ = 10$^{13}$ G pulsar (bottom).  The pulsar makes dust formation occur later but more quickly, and the parameter evolution is qualitatively similar in both cases.}
\label{fig:ourtd}
\end{figure}

In Figure~\ref{fig:ourtd}, we compare the time evolution of $\ln(S)$, $I_s$, $f_{\text{con}}$ and $a_{\text{ave}}$ for both types of dust in the Ib5-1 composition without a pulsar (top) and with a $P$ = 2 ms, $B$ = 10$^{13}$ G pulsar (bottom).  The pulsar delays the onset of formation (which we refer to throughout this paper as the formation timescale) and decreases the time from the beginning of dust formation until $f_{\text{con}} \sim 1$ (which we refer to throughout this paper as the condensation timescale).  The formation timescale is increased from $\sim$ 180 to $\sim$ 590 days for C dust, and from $\sim$ 215 to $\sim$ 800 days for MgSiO$_3$ dust, and the condensation timescale is decreased from $\sim$ 5 days to less than 1 day for both types of dust.  These effects are due to the slowed cooling due to the energy injection from the PWN and also the delay in dust formation due to sublimation, increasing $\ln(S)$ well above 1 when dust begins to form.  

However, the evolution of these properties is qualitatively similar to the case with no pulsar.  There is a spike in $I_s$ corresponding to the sudden nucleation of dust throughout the ejecta, which causes $f_{\text{con}}$ to jump to $\sim$ 1 within the condensation timescale.  After this, the supersaturation ratio drops because of the drop in gas concentration, and this causes $I_s$ to fall to $\sim$ 0.  As time goes on, the nucleated grains grow in size by accreting free key molecules, causing the gas concentration to drop further, but more slowly than during nucleation.  There is also nucleation of new grains as the temperature drops further, but the growth rate is small and concentration evolution is dominated by growth of previously nucleated grains.  However, the growth of these grains is very slow once $f_{\text{con}} \sim 1$, as $a_{\text{ave}}$ stays relatively constant after this time.  We see $\ln(S)$ fall and then rise again at later times; this second rise corresponds to the point when grain growth drops off.  These results are also qualitatively similar to the high density case from \cite{2013ApJ...776...24N}, which is where the steady state formulation agreed with the more rigorous non-steady-state model, although our condensation timescale is shorter due to our ejecta being confined to a smaller volume, thus being even more dense.

\subsection{Formation Timescale and Ionization}

\begin{figure}
\begin{subfigure}
  \centering
  \includegraphics[width=\linewidth]{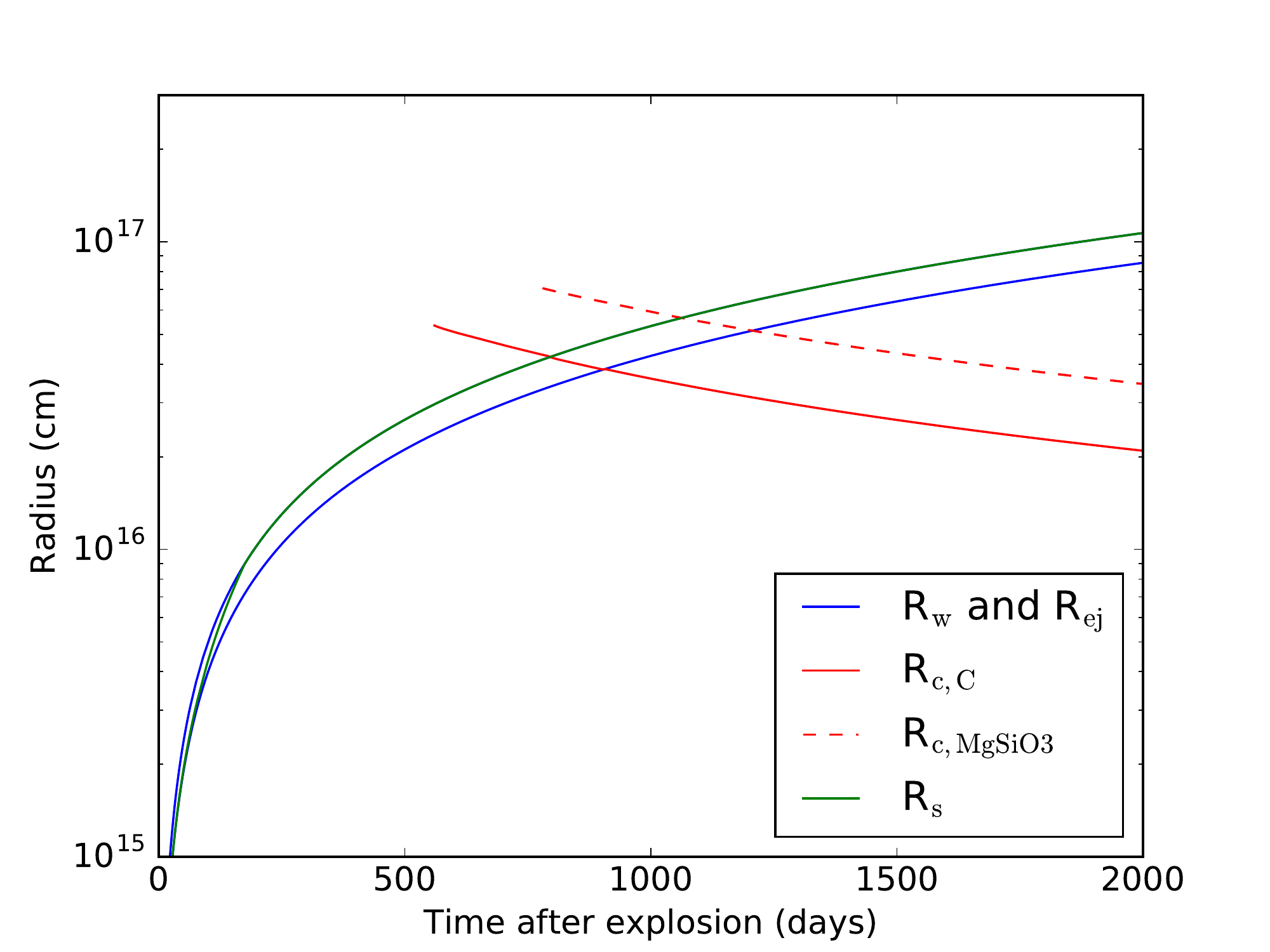}
\end{subfigure} \\
\begin{subfigure}
  \centering
  \includegraphics[width=\linewidth]{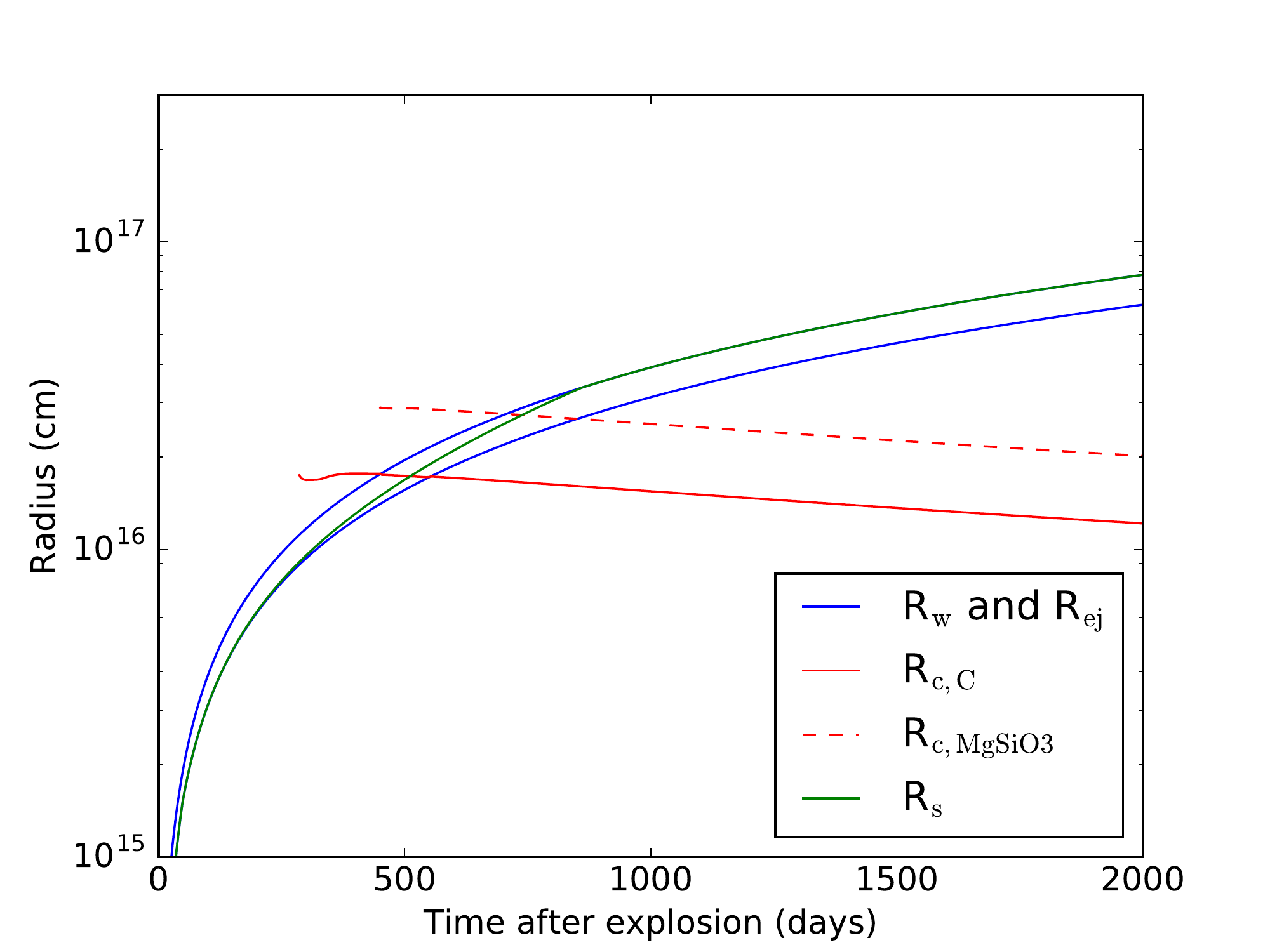}
\end{subfigure} \\
\begin{subfigure}
  \centering
  \includegraphics[width=\linewidth]{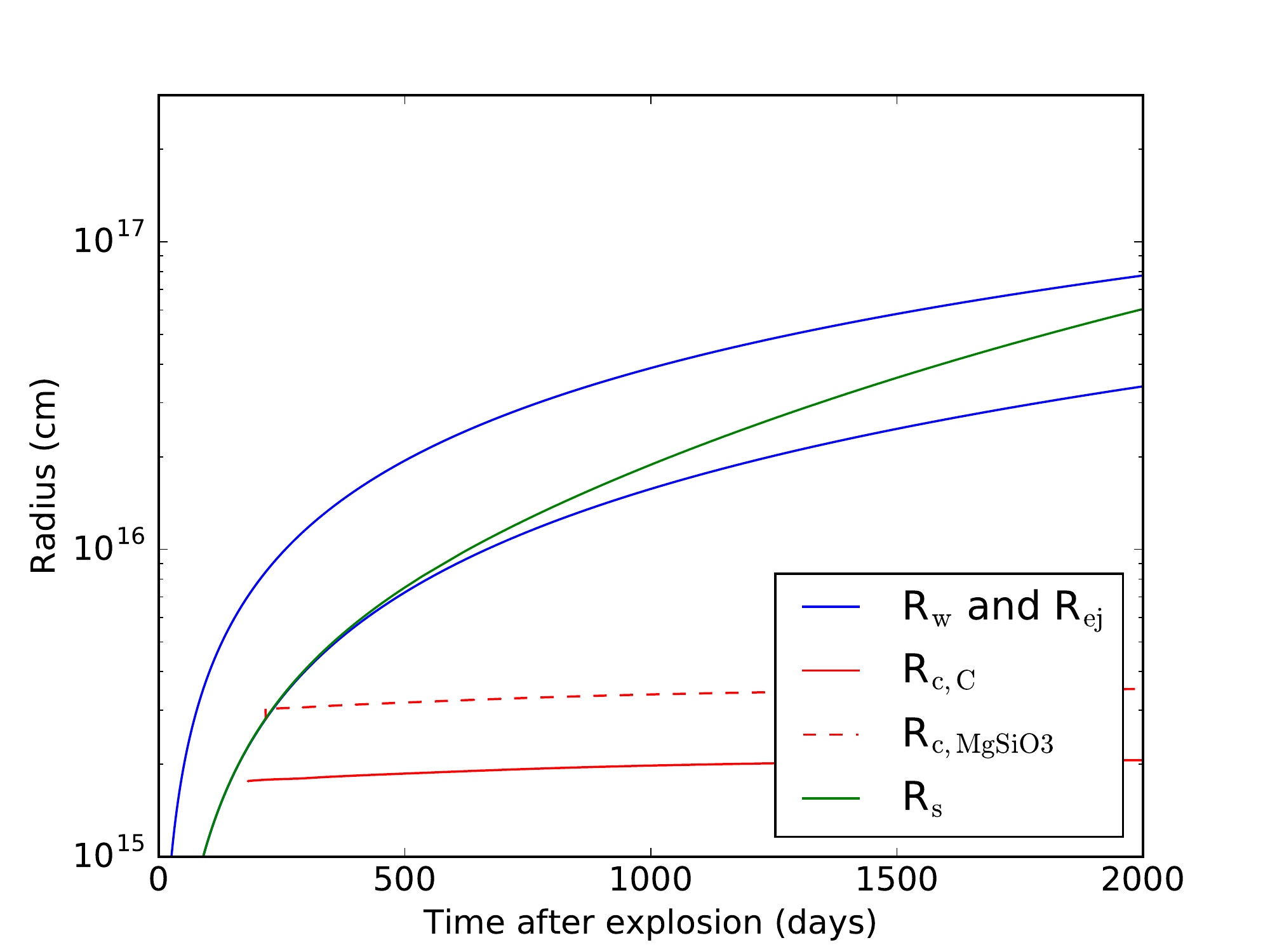}
\end{subfigure}
\caption{The time evolution of the ejecta inner $R_{\text{w}}$ and outer radius $R_{\text{ej}}$ (blue), critical (sublimation) radii $R_{\text{c}}$ for both C (solid red) and MgSiO$_3$ (dashed red) dust, and Str{\"o}mgen (ionization) radius $R_{\text{s}}$ (green), for the Ib5-1 parameters with $B$ = 5 $\times$ 10$^{12}$ G and $P$ = 1 (top), 3 (middle), and 10 ms (bottom).  The sublimation radius is shown from the point where the supersaturation ratio $S$ first becomes greater than 1.}
\label{fig:radev}
\end{figure}

In Figure~\ref{fig:radev}, we show the evolution of the ejecta inner $R_{\text{w}}$ and outer radius $R_{\text{ej}}$, critical (sublimation) radii $R_{\text{c}}$ for both types of dust, and Str{\"o}mgen (ionization) radius $R_{\text{s}}$, for the Ib5-1 parameters with initial dipole field $B$ = 5 $\times$ 10$^{12}$ G and initial rotation periods $P$ = 1, 3, and 10 ms.  The sublimation radius is shown from the point where the supersaturation ratio $S$ first becomes greater than 1, and thus $T_{\rm gas}$ becomes less than $T_c$, even if dust would be sublimated as soon as it starts to form.

The PWN emission gets stronger as $P$ decreases, and we see multiple effects because of this.  The formation time for both types of dust can increase, due to the emission increasing the temperature of the ejecta.  All the radii also increase as the emission gets stronger; $R_{\text{w}}$ and $R_{\text{ej}}$ increase because the magnetized wind from the PWN accelerates the ejecta expansion, and $R_{\text{c}}$ and $R_{\text{s}}$ both increase due to increased luminosity in the optical/UV band and above the ionization energy, respectively; the increase in $R_{\text{w}}$ can lead to enhanced adiabatic cooling, which can decrease the formation timescale.  We also see the ejecta become thicker at high $P$ due to the low acceleration of the inside edge of the ejecta.

However, the increase in $R_{\text{c}}$ and $R_{\text{s}}$ as period decreases is greater than the increase in $R_{\text{w}}$, and this leads to qualitatively different dust formation behaviour.  For $P$ = 10 ms, dust formation begins in the outer region for both dust types and in the inner region for C grains as soon as the supersaturation ratio $S$ = 1; the inside region at first has MgSiO$_3$ dust grains sublimated as soon as they begin to form, but as $R_{\text{c}}$ decreases these regions will eventually begin to form dust.  $R_{\text{s}}$ is only slightly greater than $R_{\text{w}}$, so only the very inner region will be ionized.  For $P$ = 3 ms, the sublimation radii for both types of dust are outside the edge of the ejecta when the dust first starts to form, so the dust is immediately sublimated once it becomes large enough to absorb optical/UV radiation; due to smaller dust grains having lower infrared emissivity than larger ones, they are unable to radiate heat as efficiently and are thus easier to sublimate.  As the PWN luminosity decreases, $R_{\text{c}}$ drops below $R_{\text{ej}}$ and dust begins to form near the outer edge of the ejecta.  The dust-forming region will grow larger as time passes until $R_{\text{c}}$ = $R_{\text{s}}$, at which point the gas in the inner region will remain ionized while in the outer region the dust will remain unsublimated, although maybe partially ionized.  For $P$ = 1 ms, the entire ejecta will be at least partially ionized before dust begins to form, so it seems unlikely that a large amount of dust will be able to form at all.

\begin{figure*}
\settoheight{\tempdima}{\includegraphics[width=.33\linewidth]{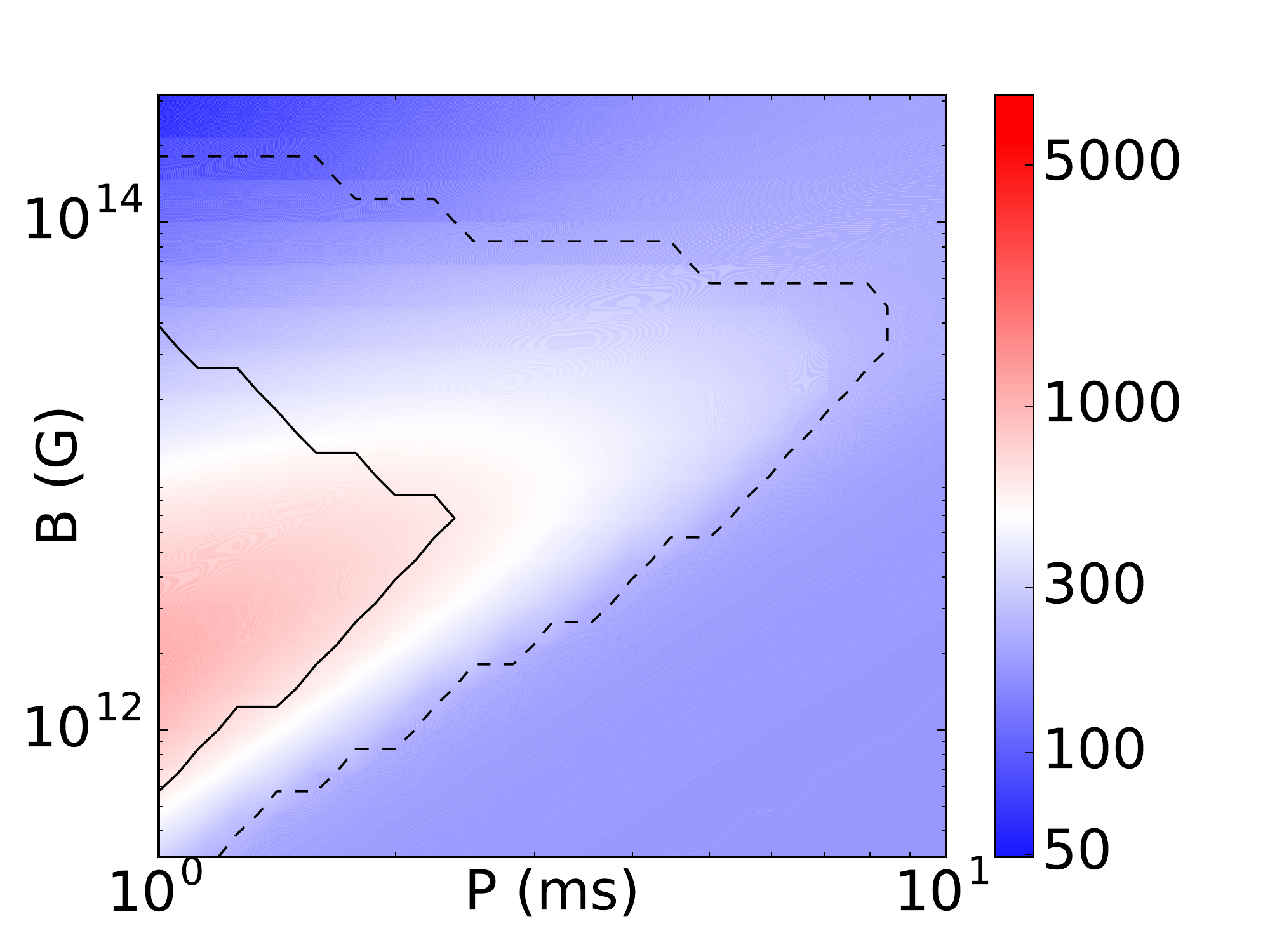}}%
\centering\begin{tabular}{@{}c@{ }c@{ }c@{}}
&\textbf{C dust} & \textbf{MgSiO$_3$/MgO dust} \\
\rowname{Ib5-1}&
\includegraphics[width=.33\linewidth]{pbdia_b15_c}&
\includegraphics[width=.33\linewidth]{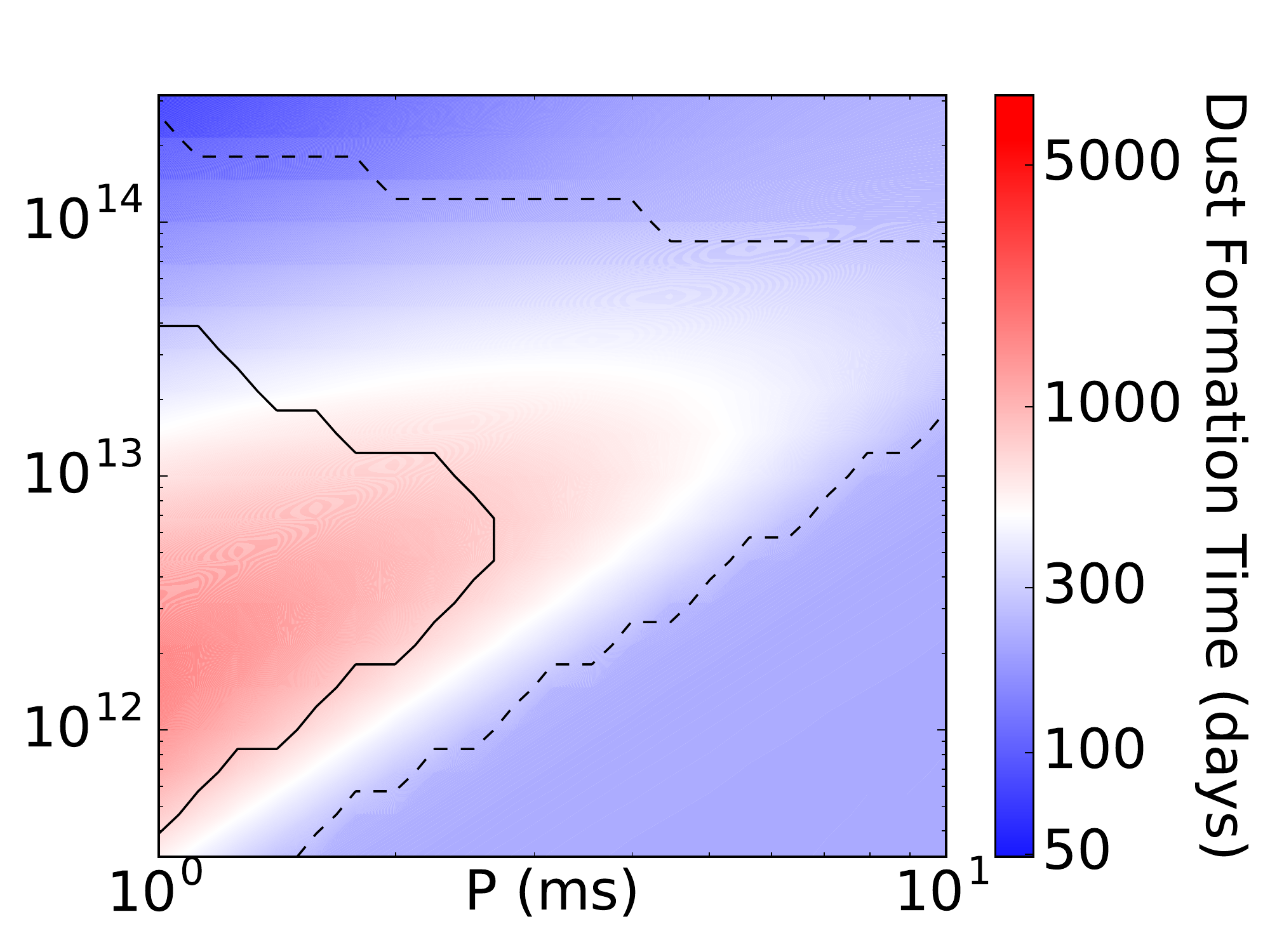}\\[-1ex]
\rowname{Ib5-05}&
\includegraphics[width=.33\linewidth]{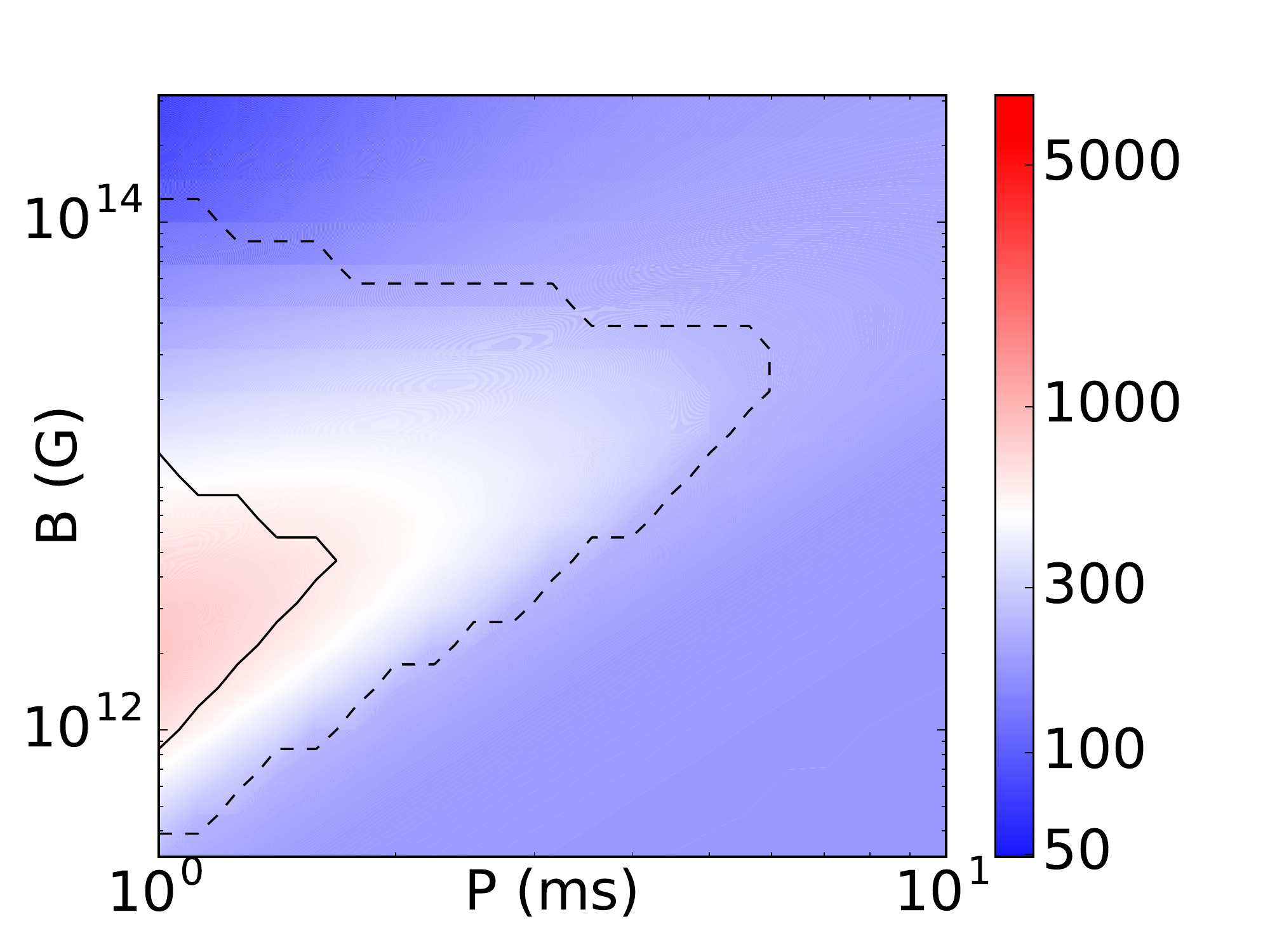}&
\includegraphics[width=.33\linewidth]{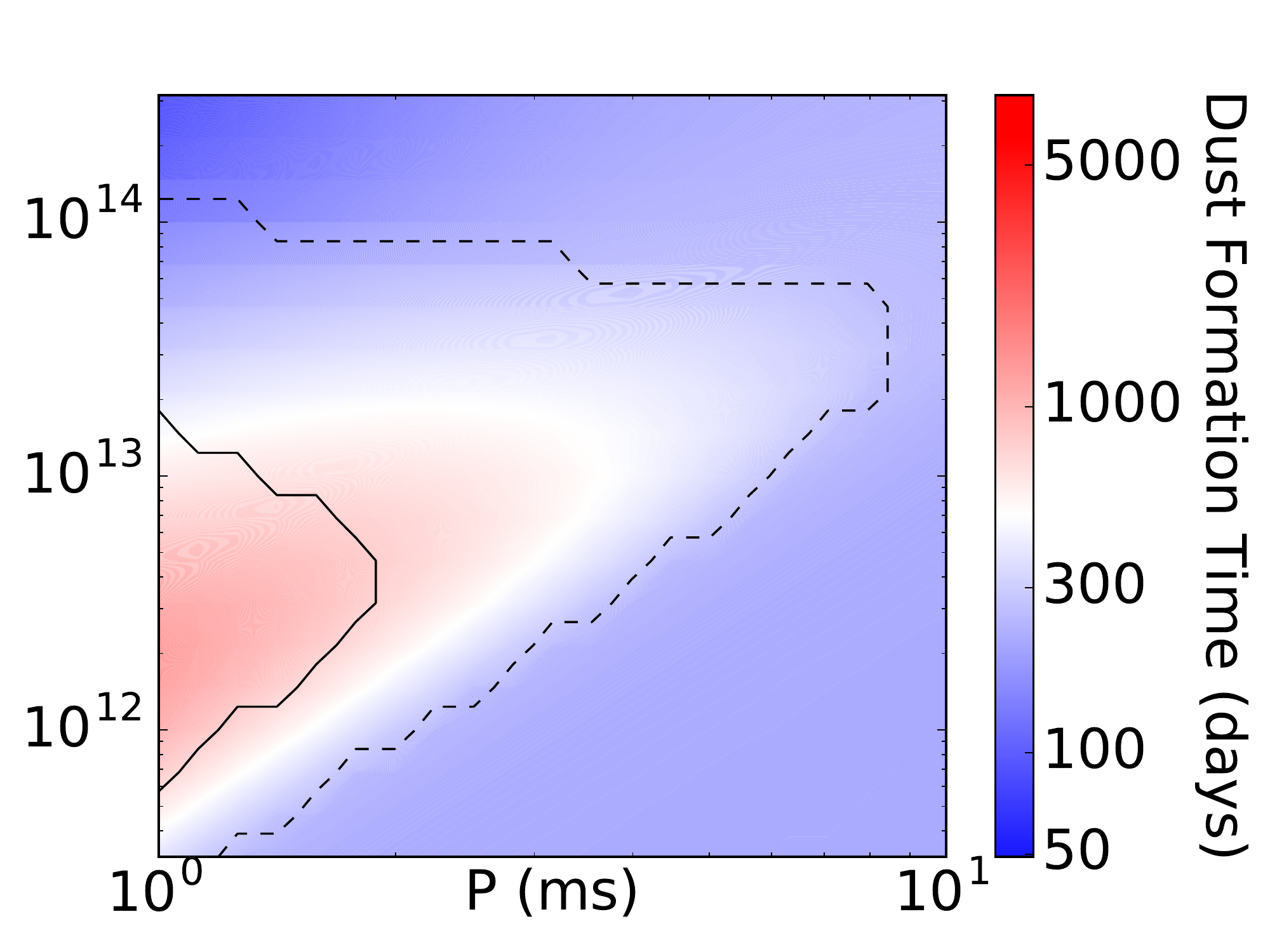}\\[-1ex]
\rowname{Ib15-1}&
\includegraphics[width=.33\linewidth]{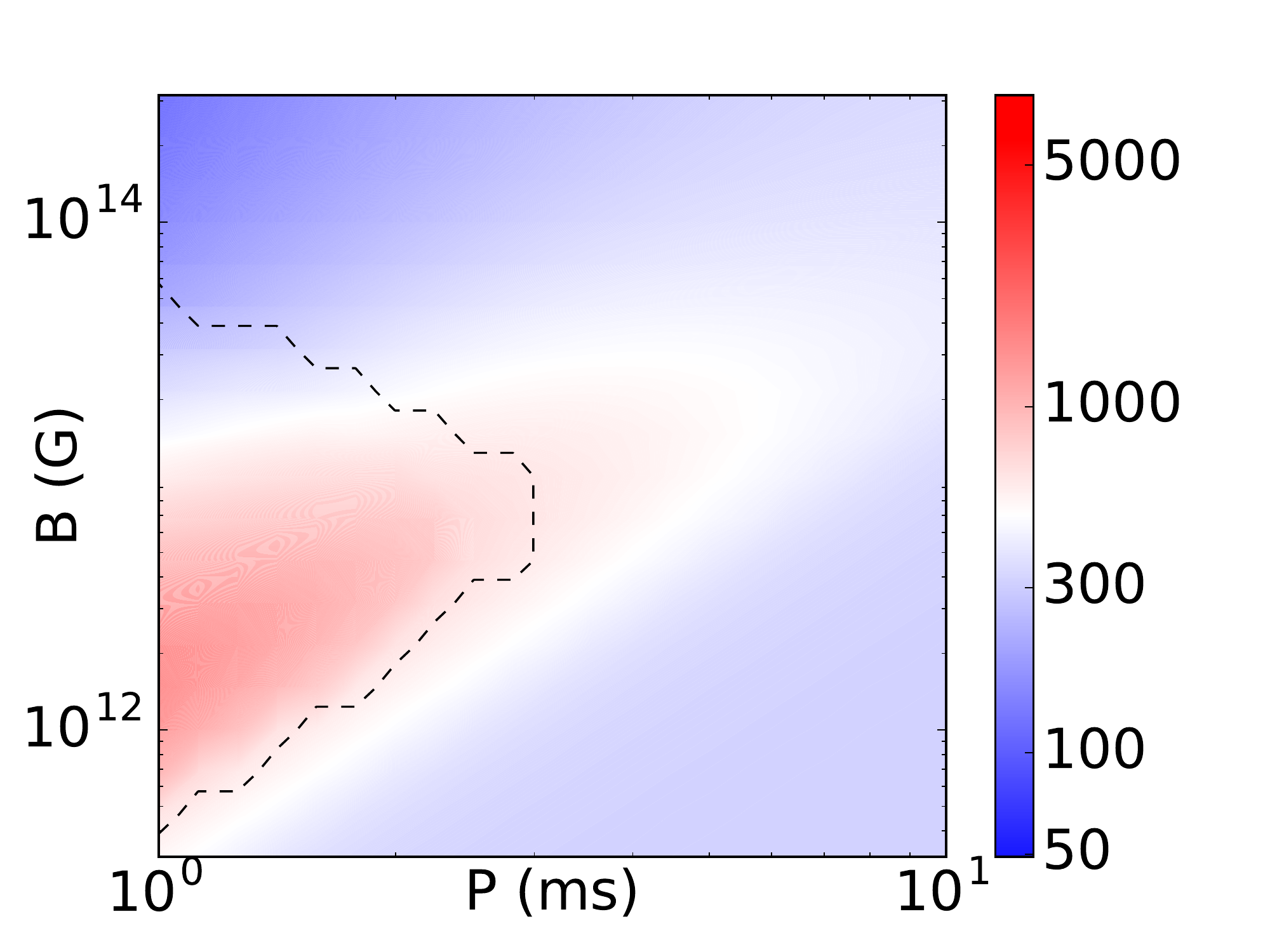}&
\includegraphics[width=.33\linewidth]{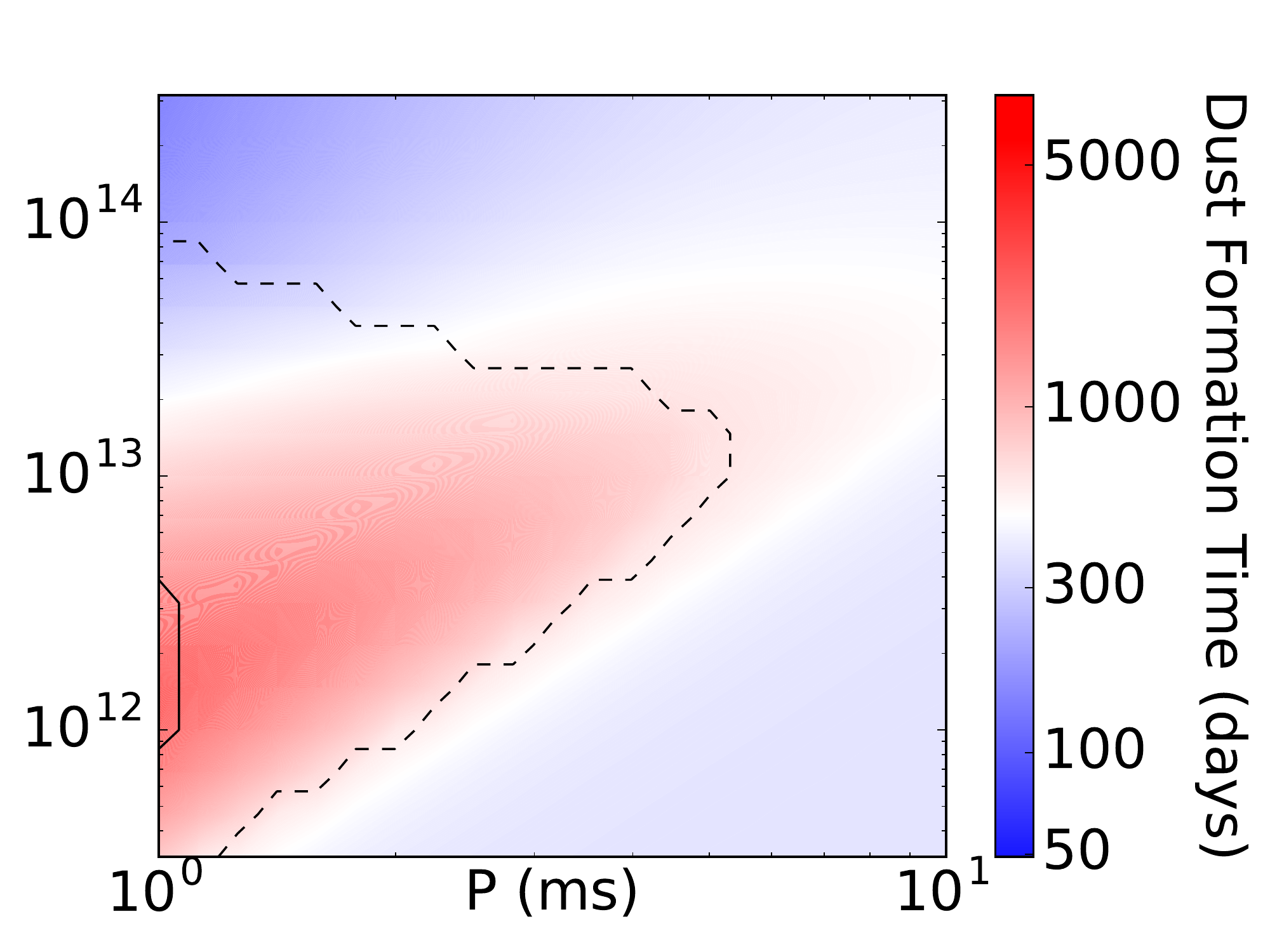}\\[-1ex]
\rowname{Ic5-1}&
\includegraphics[width=.33\linewidth]{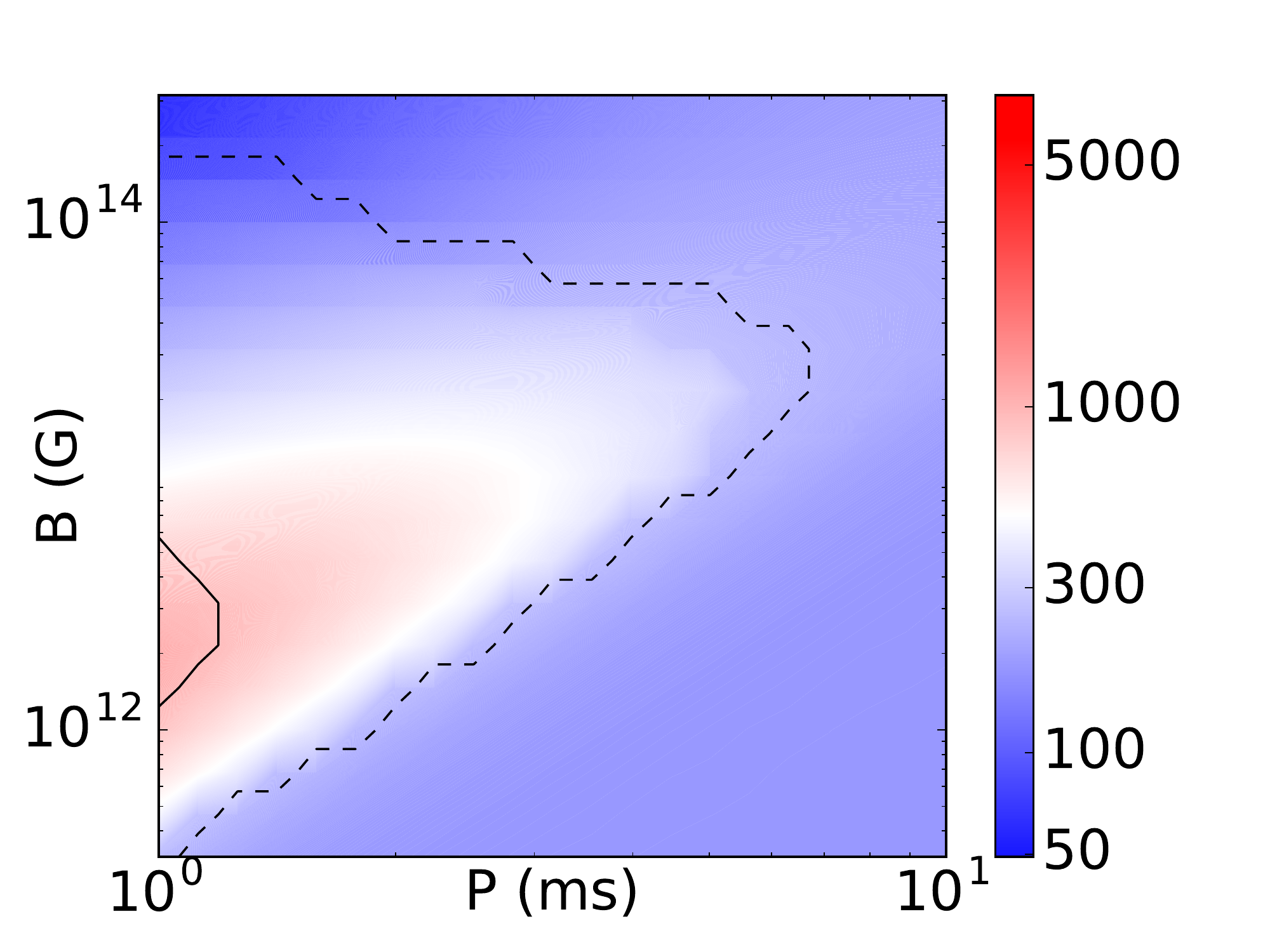}&
\includegraphics[width=.33\linewidth]{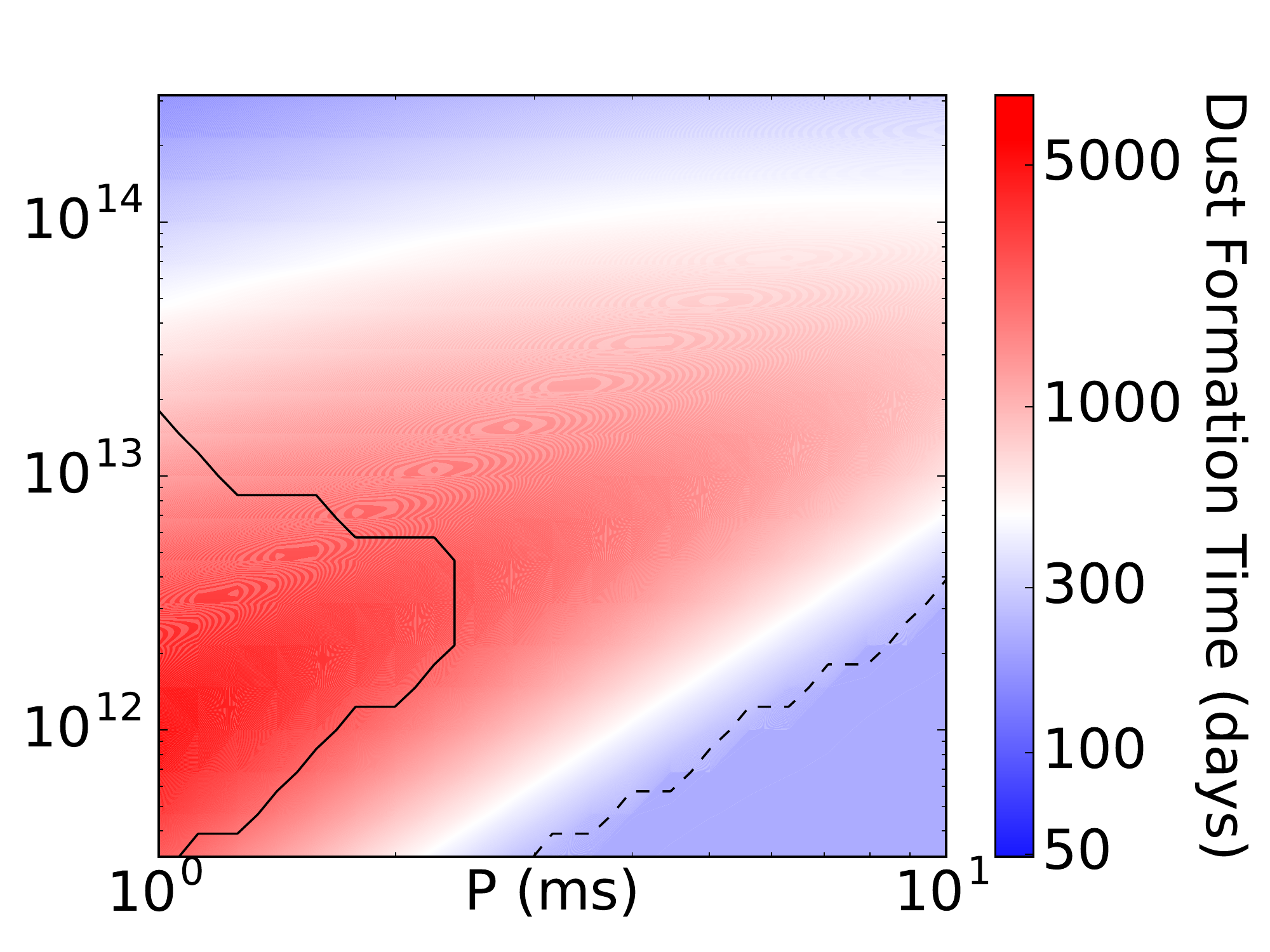}\\[-1ex]
\rowname{Ic15-1}&
\includegraphics[width=.33\linewidth]{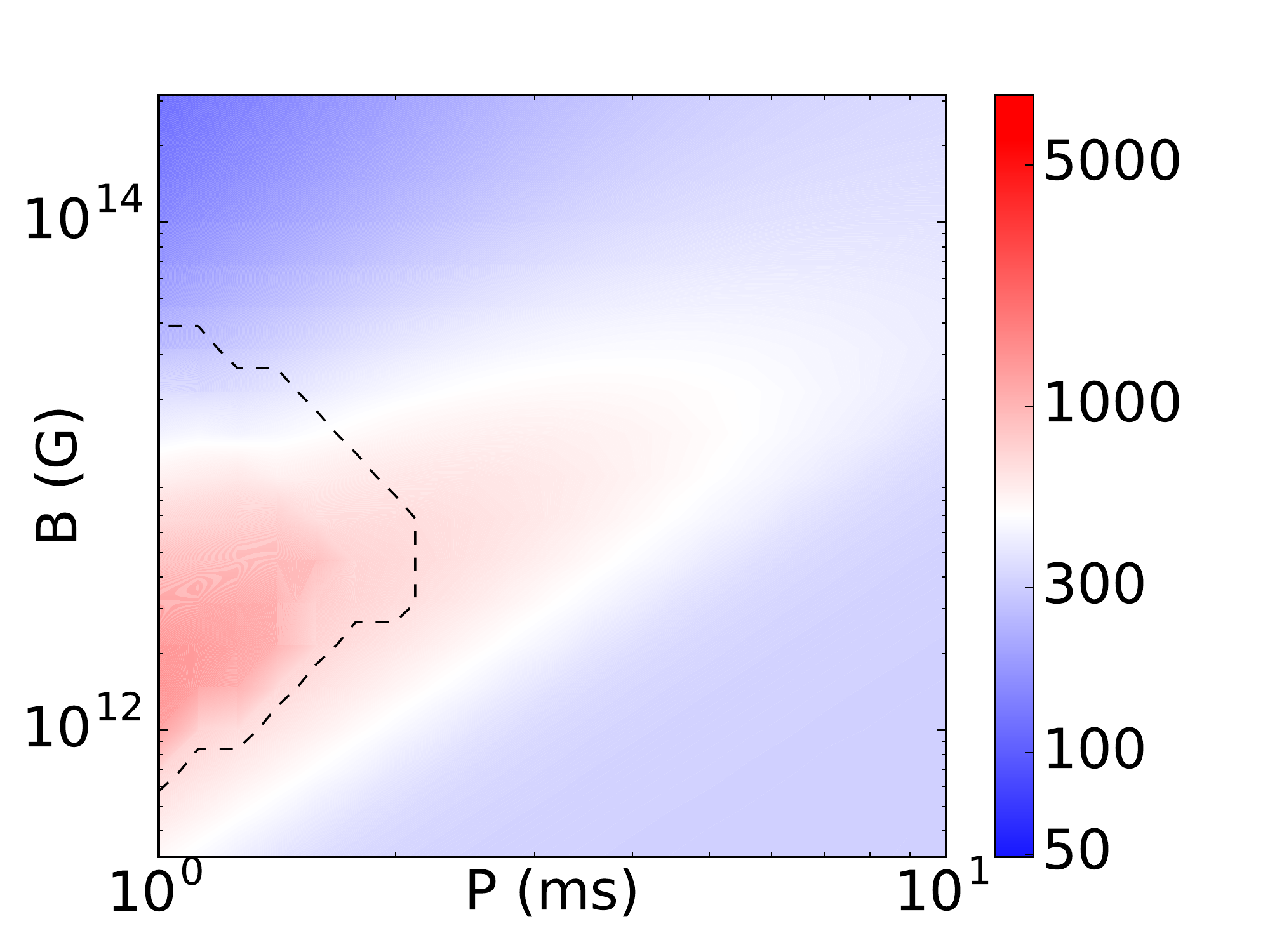}&
\includegraphics[width=.33\linewidth]{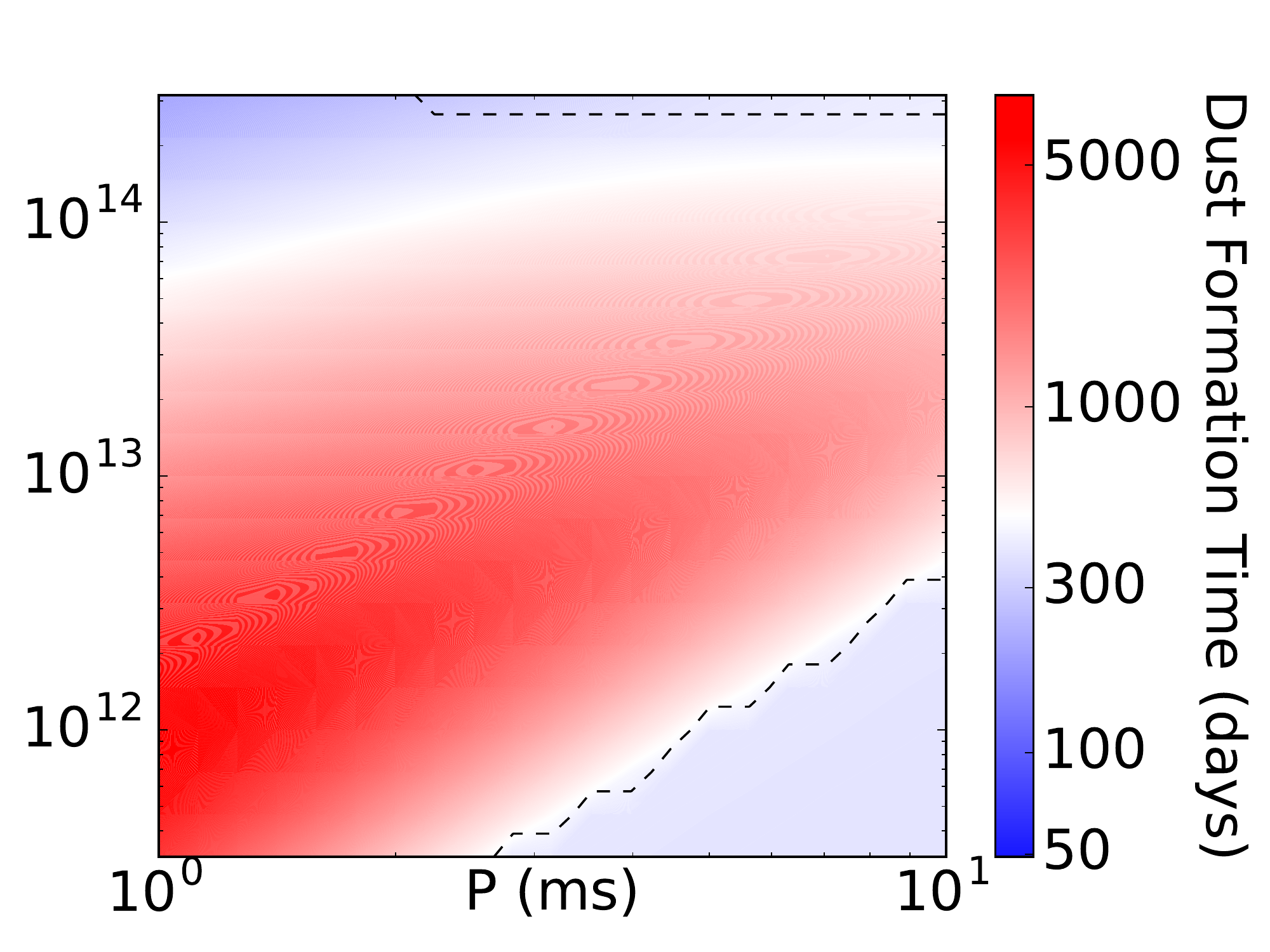}\\[-1ex]
\end{tabular}
\caption{Dependence of formation timescale for C and MgSiO$_3$ (in the Ib composition) or MgO (in the Ic composition) dust on $B$ and $P$.  The dashed black line indicates when dust formation starts to be delayed due to sublimation, and the solid black line indicates where the ejecta is fully ionized before dust formation begins, which may stop dust formation altogether (Eg. Figure \ref{fig:radev}).  Numerical values for the minimum and maximum formation times, as well as the formation time with no pulsar, are given in Table~\ref{tbl:formt}.}%
\label{fig:formt}
\end{figure*}

In Figure~\ref{fig:formt}, we show the formation timescale for C and MgSiO$_3$ (in the Ib composition) or MgO (in the Ic composition) dust for all parameters shown in Table~\ref{tbl:planrun}.  The dashed black line indicates when dust formation starts to be delayed due to sublimation, and the solid black line indicates where the ejecta is fully ionized before dust formation begins, which may stop dust formation altogether.  Numerical values for the minimum and maximum formation times, as well as the formation time with no pulsar, are given in Table~\ref{tbl:formt}.

Each graph has some qualitative features in common.  The shortest formation timescale is for low $P$ and high $B$; this is because the high initial energy injection causes the ejecta to expand very quickly, which makes adiabatic gas cooling more effective, and the fast spin-down time of the pulsar means the PWN luminosity drops very quickly, so the ejecta heating is minimal.  As the $B$ field drops, the lower ejecta velocity and slowly declining PWN emission increase the formation timescale and sublimation inside the dashed region increase it even more, like in Figure~\ref{fig:radev} (middle).  With some parameters, there is a $P$-$B$ region which has the longest formation time where the ejecta will be fully ionized before dust starts to form, like in Figure~\ref{fig:radev} (top).  In this region, it is likely that the ionization breakout will prevent dust formation entirely.  However, decreasing $B$ even further will cause the formation timescale to drop as the effects of the PWN get weaker and eventually become negligible.  As $P$ increases, the PWN also gets weaker, and can cause the formation timescale to increase, as the ejecta velocity and thus adiabatic cooling decrease, or cause the formation timescale to decrease, as the heating from the PWN decreases.  The balance between these two determines the formation timescale.

The different types of dust have different formation timescales due to both the mass fractions of their constituent gas atoms in the ejecta as well as their thermodynamic properties.  The formation timescale of MgSiO$_3$ dust is about 120\% of the timescale for C dust for accelerated or pulsar-free formation and 140-150\% for delayed formation.  The parameter space for sublimation delay and for ionization expands slightly for MgSiO$_3$ compared to C.  The factor difference between MgO formation time and C formation time vary more than with MgSiO$_3$.  For pulsar-free formation, the MgO timescale is similar to MgSiO$_3$, being around 120\% of the timescale for C.  For pulsar-accelerated formation, the difference is about 150\% for high ejecta mass and almost 250\% for low ejecta mass because formation is delayed longer by sublimation at low mass.  Pulsar-delayed formation gives the highest discrepancy though, with MgO dust taking roughly 5 times longer to form than C.  The parameter space for sublimation delay and for ionization expands significantly for MgO compared to C, with a significant amount of the parameter space we examined delayed by sublimation.

The parameter sets have many quantitative differences due to the effects of mass and luminosity on expansion and energy injection. Decreasing the PWN luminosity (compare Ib5-1 and Ib5-05) decreases the ejecta acceleration, thermal energy injection, and non-thermal ionization and sublimation.  This decreases the maximum formation timescale and increases the minimum timescale, bringing everything closer to the pulsar-free scenario, which is the limit of decreased PWN luminosity.  However, even though the luminosity was cut by 50\%, the formation timescales only changed by 20-30\% and the parameter space for sublimation delay and ionization do not change very much.  Increasing the mass (compare Ib5-1 and Ib15-1, and Ic5-1 and Ic15-1) slows the expansion of the ejecta, slowing adiabatic cooling and and increasing the energy flux from the PWN, heating up the ejecta.  As a result, the formation timescale is increased for all scenarios, varying from increases of around 30\% for delayed formation to 100\% for accelerated formation, except for MgO, which only increases by around 20\% for accelerated formation.

\begin{table*}
\begin{tabular}{|c|ccc|ccc|ccc|} \hline
 & \multicolumn{3}{c}{C dust}&\multicolumn{3}{c}{MgSiO$_3$ dust}&\multicolumn{3}{c}{MgO dust}\\
ID & $t_{\text{max}}$ & $t_{\text{min}}$ & $t_{\text{no pulsar}}$ & $t_{\text{max}}$ & $t_{\text{min}}$ & $t_{\text{no pulsar}}$ & $t_{\text{max}}$ & $t_{\text{min}}$ & $t_{\text{no pulsar}}$ \\ \hline
Ib5-1 & 1118 & 58 & 180 & 1649 & 73 & 215 &&& \\
Ib5-05 & 883 & 72 & 180 & 1263 & 88 & 215 &&& \\
Ib15-1 & 1498 & 120 & 316 & 2180 & 141 & 376 &&&  \\
Ic5-1 & 1072 & 58 & 177 &&&& 5030 & 143 & 216 \\ 
Ic15-1 & 1420 & 118 & 311 &&&& 6657 & 175 & 378 \\ \hline 
\end{tabular}
\caption{Numerical values in days for the minimum and maximum formation times, as well as the formation time with no pulsar, for all parameters shown in Table~\ref{tbl:planrun}.  The formation timescale dependence on $B$ and $P$ is shown in Figure~\ref{fig:formt}.}
\label{tbl:formt}
\end{table*}

\subsection{Dust Size Distribution}

\begin{figure*}
\settoheight{\tempdima}{\includegraphics[width=.33\linewidth]{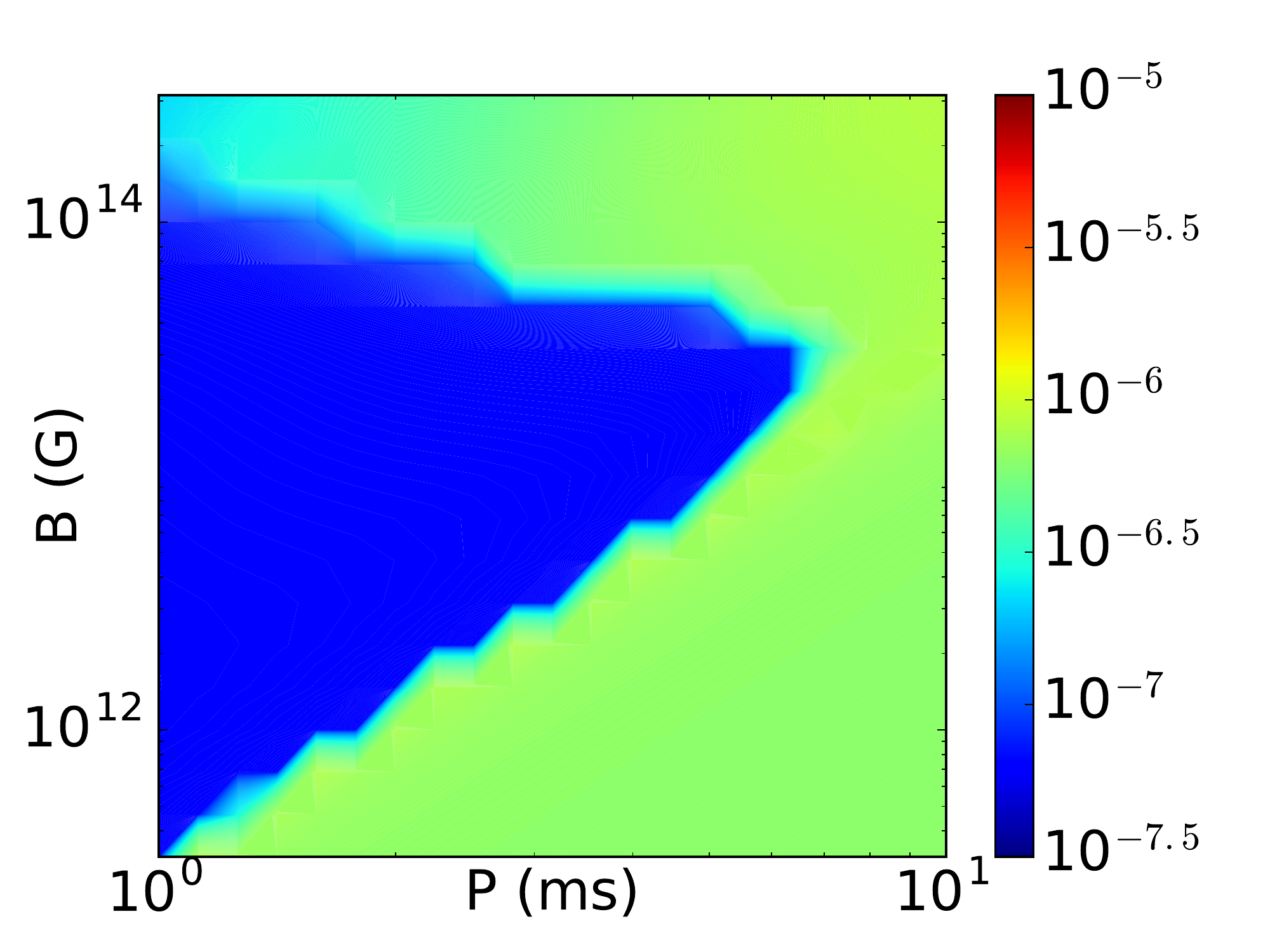}}%
\centering\begin{tabular}{@{}c@{ }c@{ }c@{}}
&\textbf{C dust} & \textbf{MgSiO$_3$/MgO dust} \\
\rowname{Ib5-1}&
\includegraphics[width=.33\linewidth]{pbdia_b15_c_adist}&
\includegraphics[width=.33\linewidth]{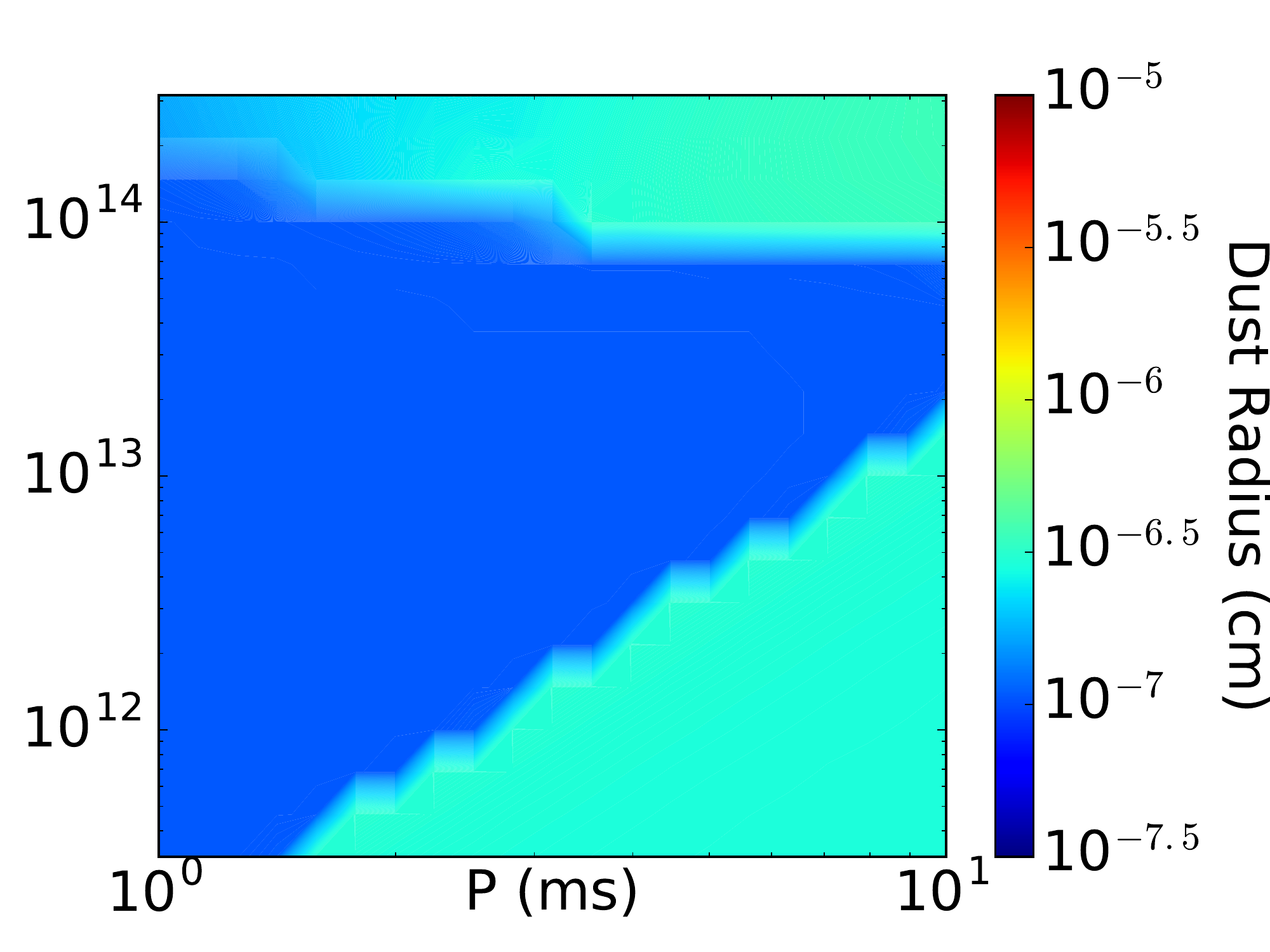}\\[-1ex]
\rowname{Ib5-05}&
\includegraphics[width=.33\linewidth]{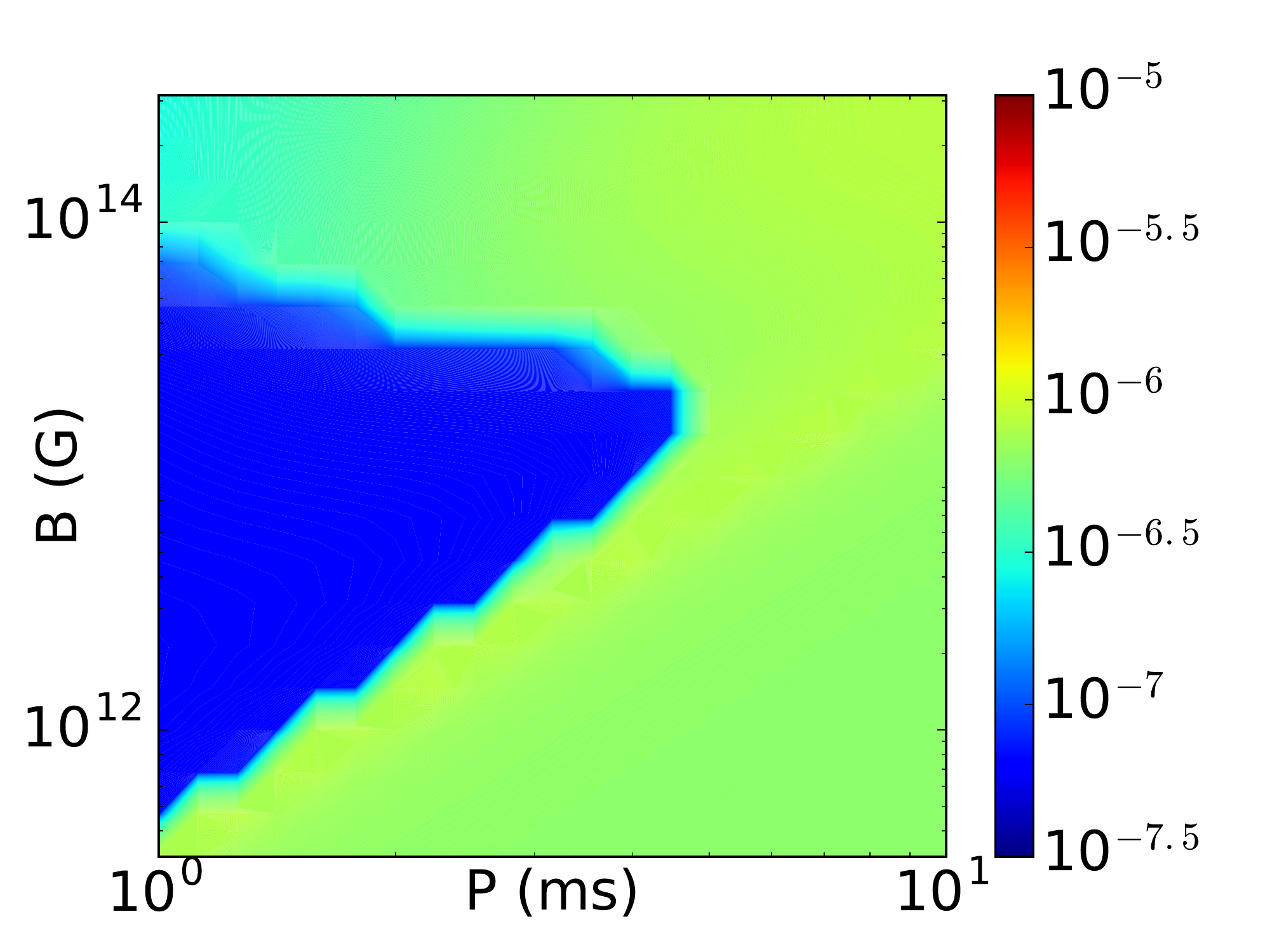}&
\includegraphics[width=.33\linewidth]{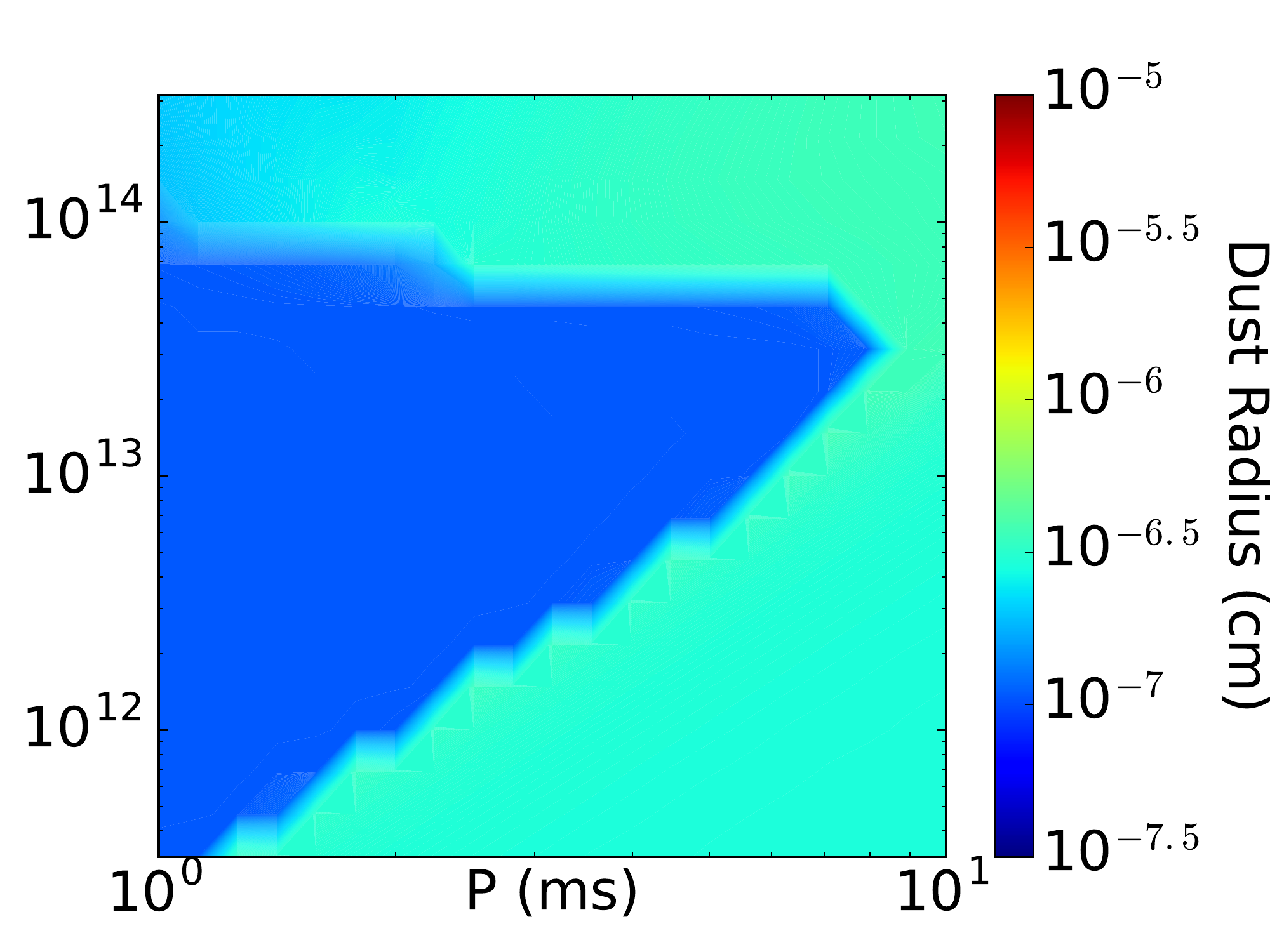}\\[-1ex]
\rowname{Ib15-1}&
\includegraphics[width=.33\linewidth]{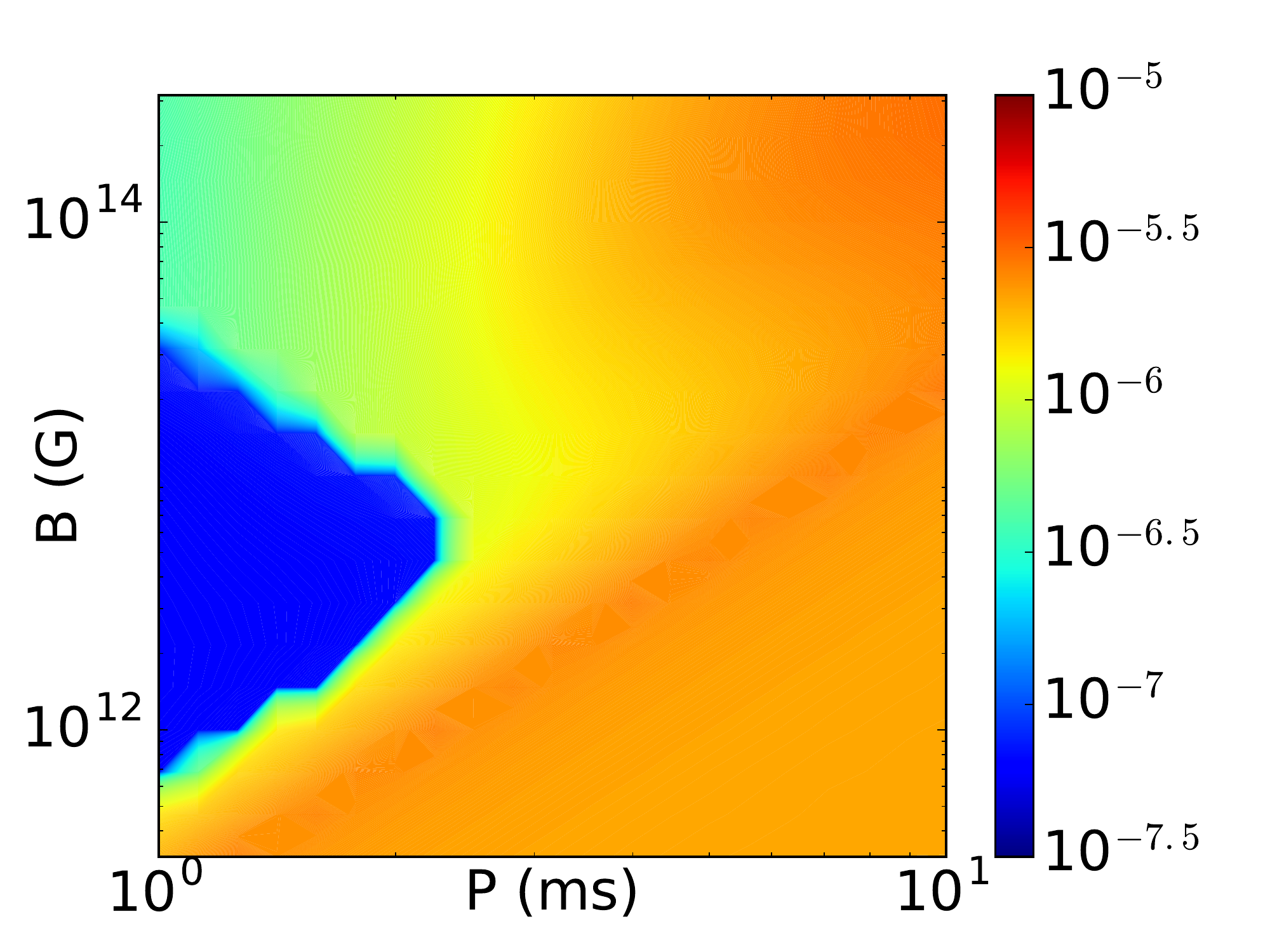}&
\includegraphics[width=.33\linewidth]{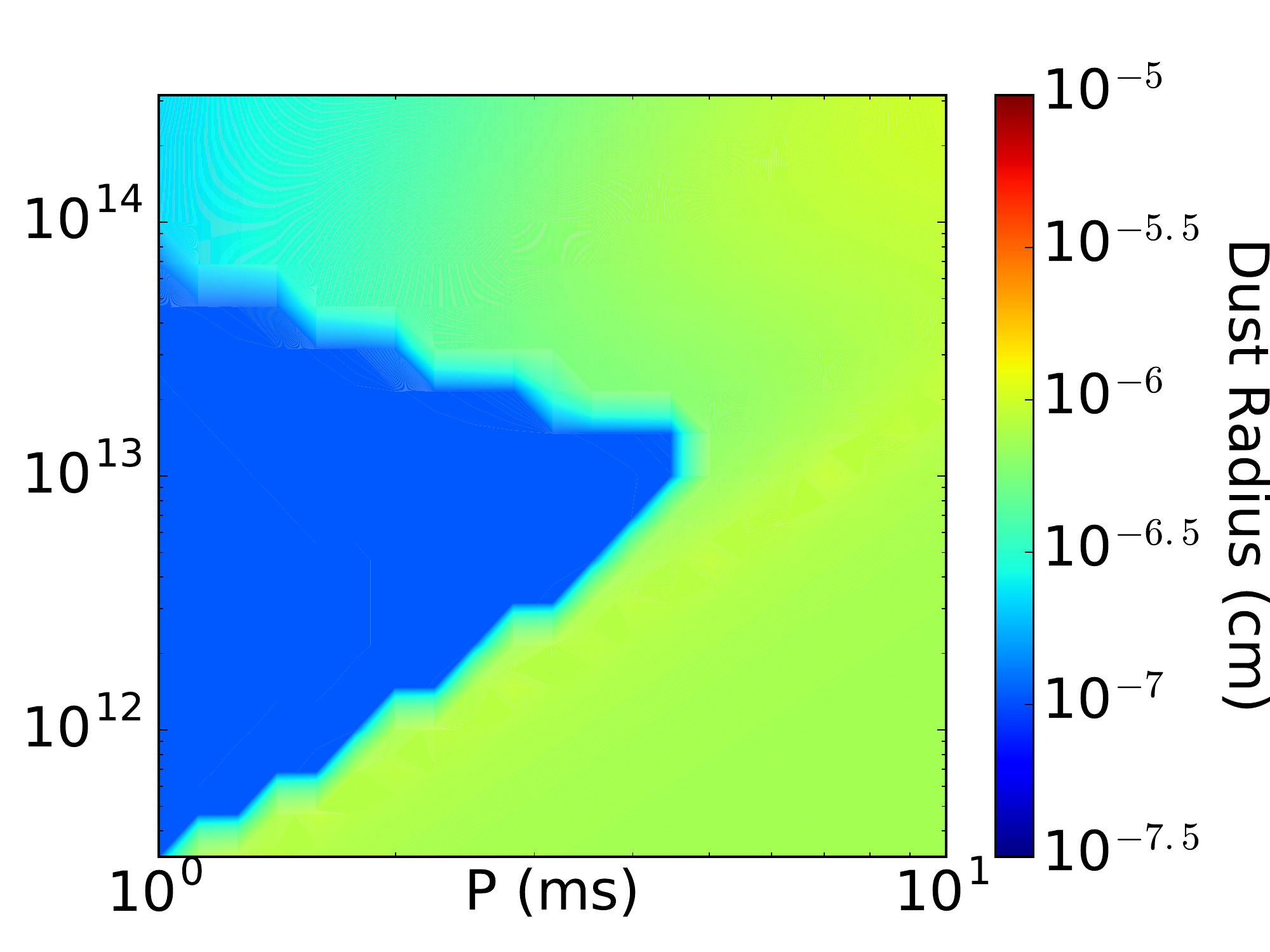}\\[-1ex]
\rowname{Ic5-1}&
\includegraphics[width=.33\linewidth]{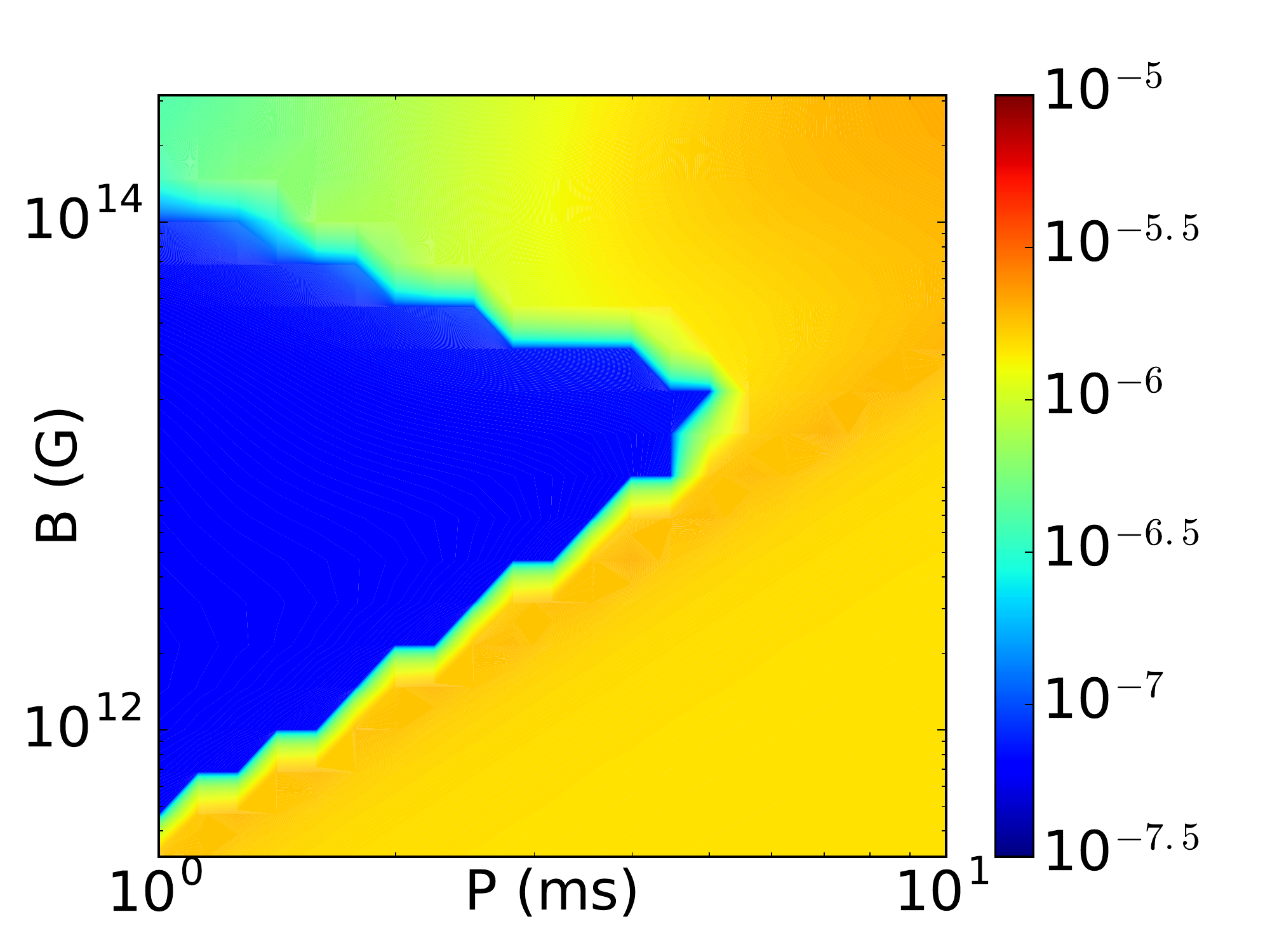}&
\includegraphics[width=.33\linewidth]{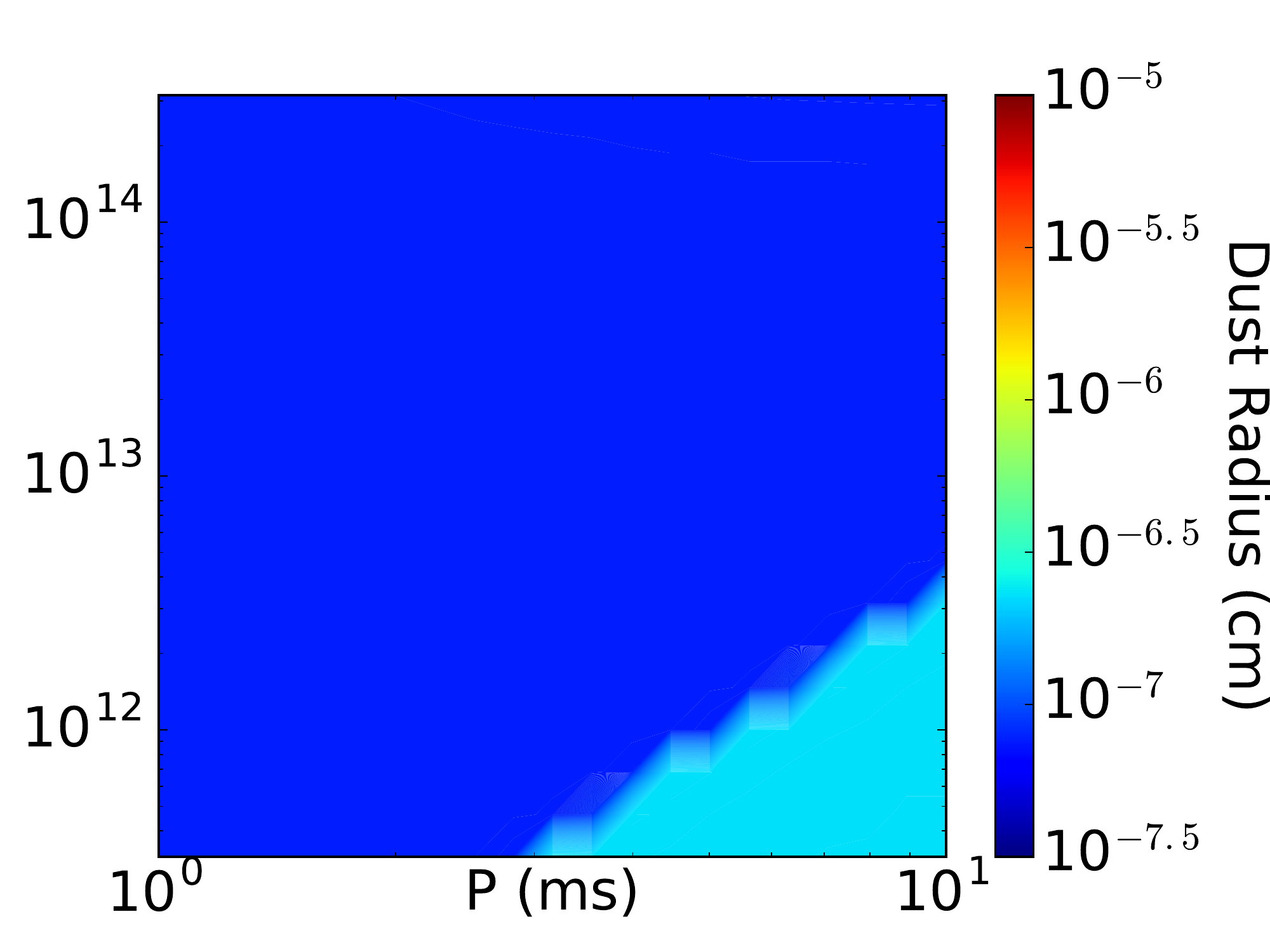}\\[-1ex]
\rowname{Ic15-1}&
\includegraphics[width=.33\linewidth]{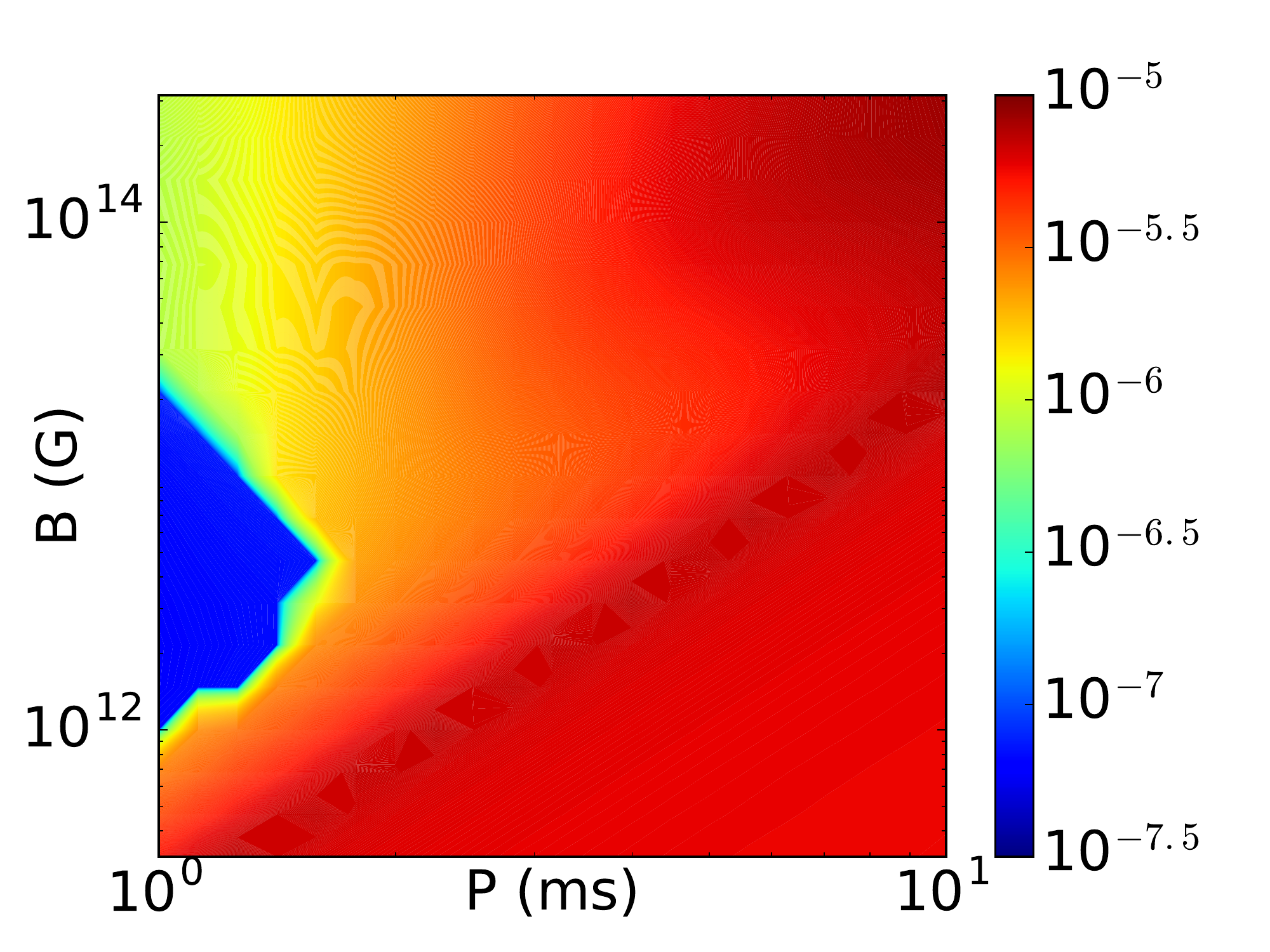}&
\includegraphics[width=.33\linewidth]{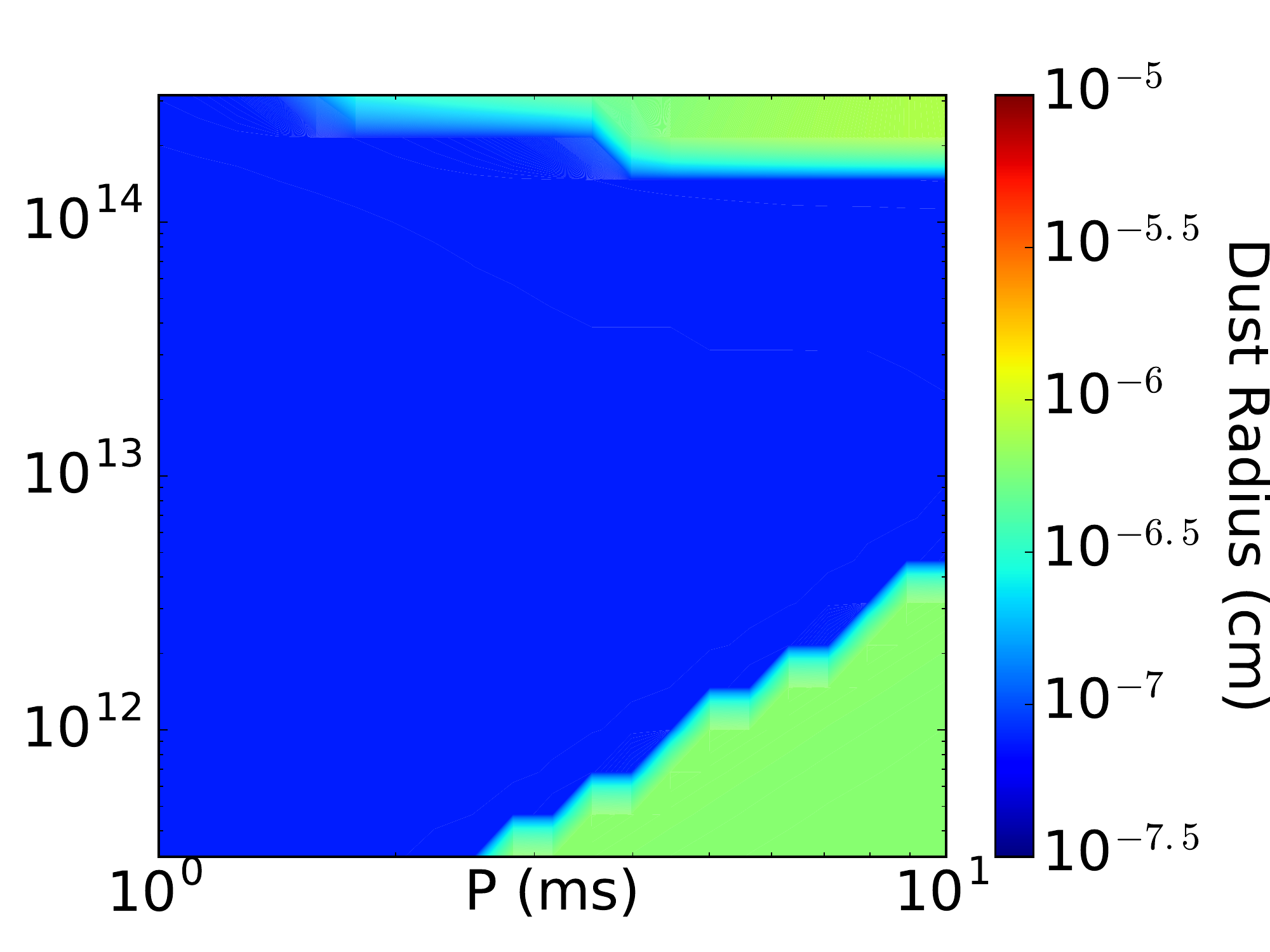}\\[-1ex]
\end{tabular}
\caption{Dependence of final average dust size for C and MgSiO$_3$ (in the Ib composition) or MgO (in the Ic composition) dust on $B$ and $P$.}%
\label{fig:adist}
\end{figure*}

In Figure~\ref{fig:adist}, we show the final average size distribution for C and MgSiO$_3$ (in the Ib composition) or MgO (in the Ic composition) dust for all parameter sets shown in Table~\ref{tbl:planrun}, and show the minimum and maximum size for each in Table \ref{tbl:forma}.  The size of the dust is heavily dependent on the gas concentration at the time of formation.  Thus, in parameter regions where dust formation is delayed by sublimation, or the ejecta is accelerated by the PWN, the dust size is significantly smaller.  Dust size is largest when the effect of the PWN on the ejecta becomes weaker or negligible, which is why the large $P$ and small $B$ region produces the largest dust; the spin-down timescale of this neutron star is over 10$^4$ years.  Increasing ejecta mass (compare Ib5-1 and Ib15-1, and Ic5-1 and Ic15-1) and gas concentration (compare C dust in Ib5-1 and Ic5-1, and Ib15-1 and Ic15-1) also increases dust size.

\begin{table}
\begin{tabular}{|c|cc|cc|cc|} \hline
 & \multicolumn{2}{c}{C dust}&\multicolumn{2}{c}{MgSiO$_3$ dust}&\multicolumn{2}{c}{MgO dust}\\
ID & $a_{\text{max}}$ & $a_{\text{min}}$ & $a_{\text{max}}$ & $a_{\text{min}}$ & $a_{\text{max}}$ & $a_{\text{min}}$ \\ \hline
Ib5-1 & 8.4 & 0.6 & 3.6 & 1.1 && \\
Ib5-05 & 8.7 & 0.6 & 3.7 & 1.1 && \\
Ib15-1 & 30 & 0.6 & 10.0 & 1.1 &&  \\
Ic5-1 & 21 & 0.6 &&& 2.3 & 0.8 \\ 
Ic15-1 & 85 & 0.7 &&& 8.1 & 0.8 \\ \hline 
\end{tabular}
\caption{Numerical values in nm (10$^{-7}$ cm) for the minimum and maximum final average dust size, for all parameters shown in Table~\ref{tbl:planrun}.  The size distribution dependence on $B$ and $P$ is shown in Figure~\ref{fig:adist}.}
\label{tbl:forma}
\end{table}

Numerical simulations \citep{2010ApJ...715.1575S, 2012ApJ...748...12S} suggest that grains below 100 nm will be almost completely destroyed by the SN reverse shock, and larger ones will be sputtered to a smaller size \citep{2010ApJ...713..356N}.  With this criterion, most dust will be destroyed by the reverse shock, since the average dust radius is always lower than 100 nm.  However, since the dust distribution found by \cite{2013ApJ...776...24N} spans about one order of magnitude, it's possible that the largest C dust in non-pulsar-driven Type Ic SNe, or SNe with large ejecta masses, may survive the reverse shock, but the presence of a pulsar wind nebula increases the likelihood that most of the dust will be destroyed. Silicates will always be destroyed by the reverse shock unless the ejecta mass is higher than we model, as will carbon in the Ib5-1 and Ib5-05, regardless of the presence of a pulsar or not.  It is therefore unlikely that pulsar-driven SN will contribute greatly to the overall dust concentration in the ISM.

It's worth noting that the average dust size is almost constant over the entire area where formation is delayed due to sublimation (compare to the dashed lines in Figure \ref{fig:formt}); this is likely due to our use of the steady-state approximation.  This size is close to the minimum size for dust to be considered a grain, and indicates that dust can not grow far beyond the point where is can efficiently absorb continuum energy, and that the final size of this dust actually depends on detailed microphysics beyond the scope of this paper.

\subsection{Dust Emission} \label{sec:resdustem}

We are interested in the possible detection of dust emission in Type-Ic SLSN remnants, so we examine the emission for two fiducial parameter sets: the P1 set, with $P$ = 1 ms, $B$ = 10$^{14}$ G, and $M_{\rm ej}$ = 15 M$_{\sun}$; and the P2 set, with $P$ = 2 ms, $B$ = 2 $\times$ 10$^{13}$ G, and $M_{\rm ej}$ = 5 M$_{\sun}$.  These are chosen to roughly match the $P_{\rm min}$ and $M_{\rm max}$ cases from \cite{2018MNRAS.474..573O}, and both have the Ic composition.  

\begin{figure}
\begin{subfigure}
  \centering
  \includegraphics[width=\linewidth]{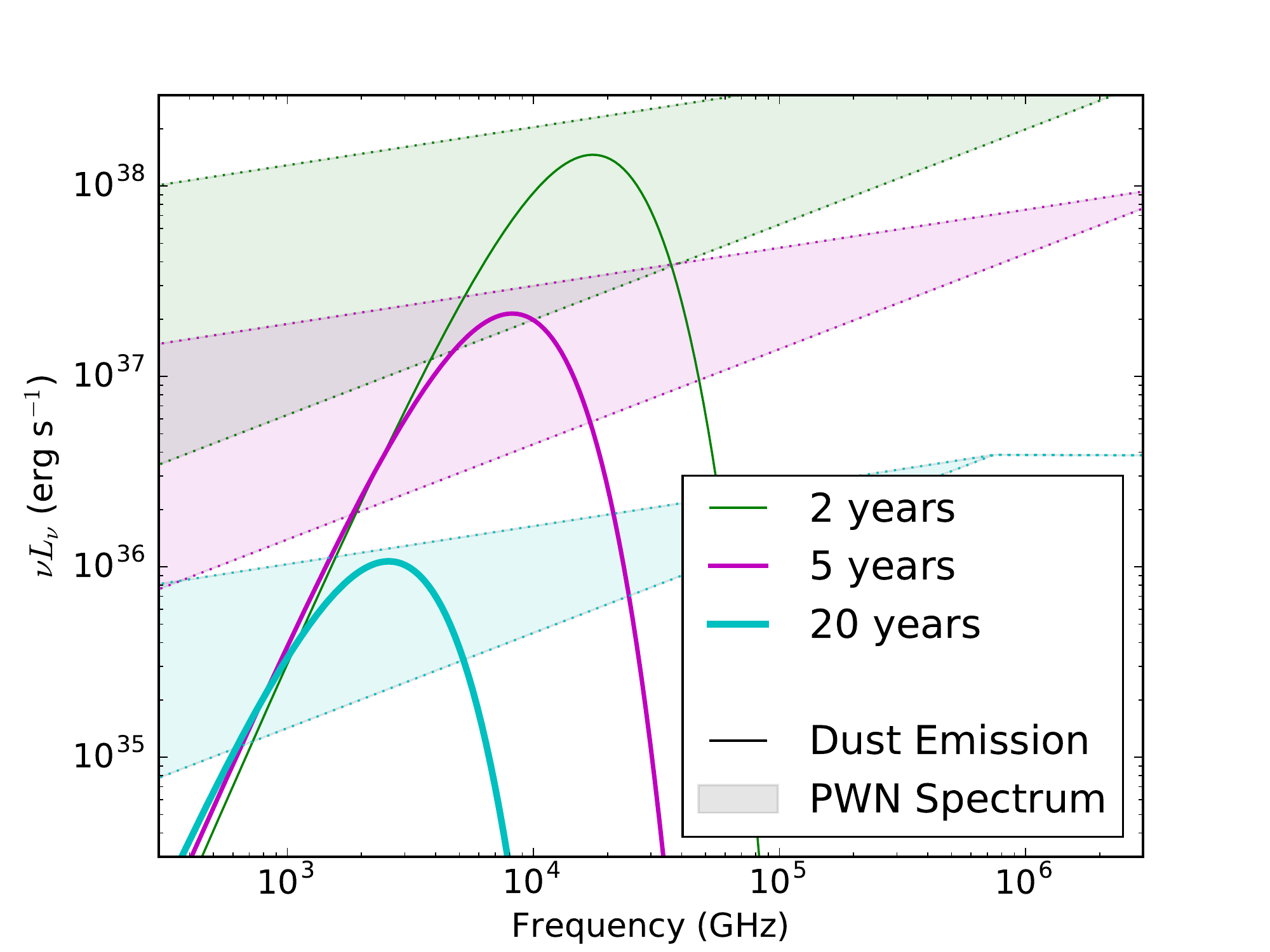}
\end{subfigure} \\
\begin{subfigure}
  \centering
  \includegraphics[width=\linewidth]{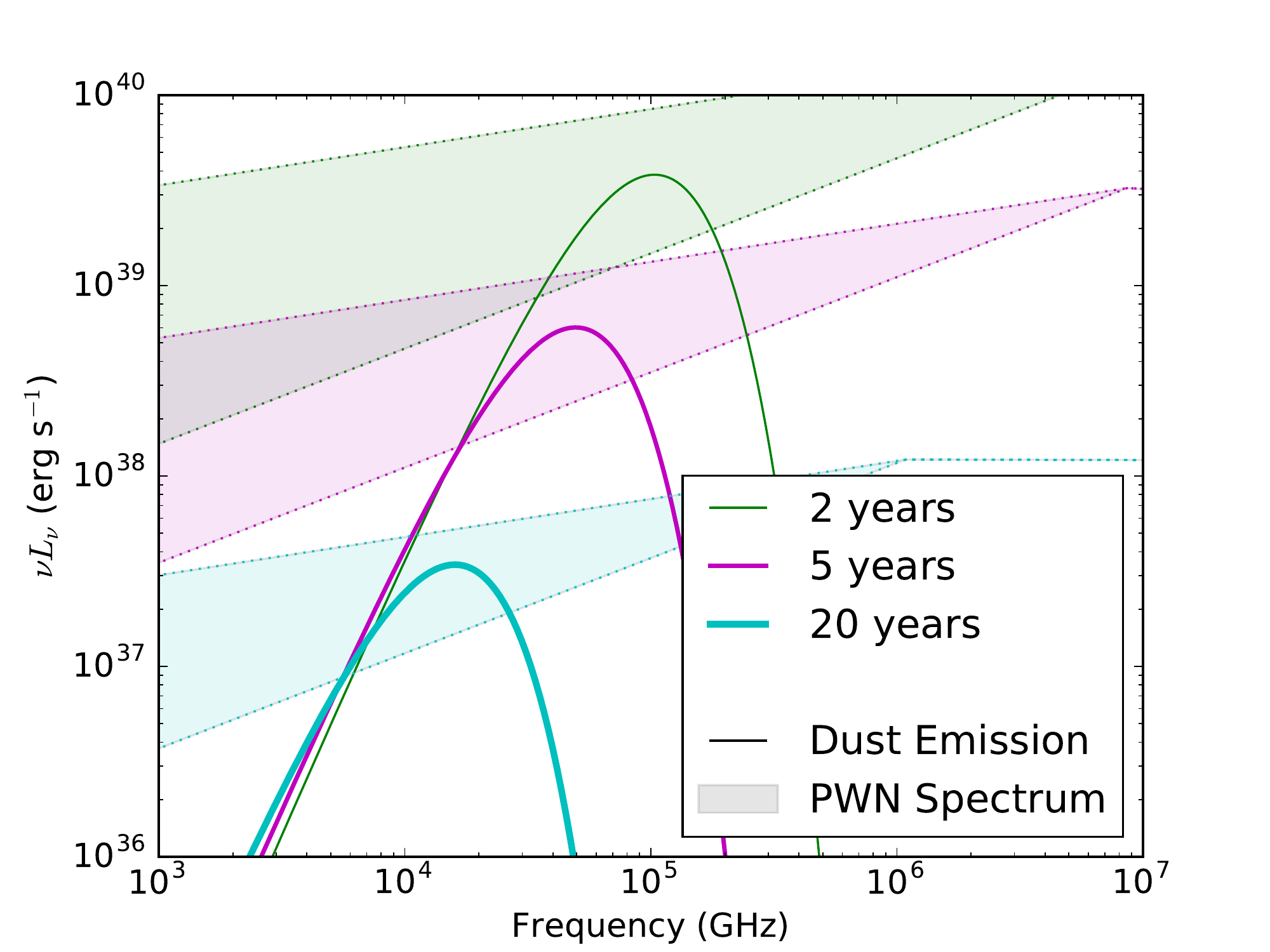}
\end{subfigure}
\caption{The PWN (dotted/shaded) and dust (dotted) spectra 2 (green), 5 (magenta), and 20 (cyan) years after the explosion.  The P1 case is shown above and the P2 case below.  The dotted lines represent the PWN spectra with $\alpha_1$ = 1.5 and 1.8, with the shaded region in between.  The dust emission was calculated with $\alpha_1$ = 1.8, and is expected to be lower by a factor of $\lesssim$ 2 in the first decade after the explosion when calculated with $\alpha_1$ = 1.5.}
\label{fig:slsn}
\end{figure}

The spectra of the PWN and dust and compared for the two cases in Figure~\ref{fig:slsn}.  We account for uncertainty in the PWN spectra by showing the region for $\alpha_1$ between 1.8 and 1.5, and discuss this spectral uncertainty further in Appendix \ref{sec:disrealspec}.  The dust spectrum shown was calculated with $\alpha_1$ = 1.8, but is expected to be lower by a factor of $\lesssim$ 2 in the first decade after the explosion when calculated with $\alpha_1$ = 1.5. We see that the detectability of the dust emission depends heavily on the spectral index, ranging from undetectable if 
$\alpha_1$ = 1.8 to easily detectable after 2 years in both cases if $\alpha_1$ = 1.5.  

For the $\alpha_1$ = 1.8 case (in which the non-thermal flux is likely to be overestimated), although the dust spectrum approaches the PWN spectrum as time passes, due to the lower absorbed energy giving the dust spectrum a lower peak frequency, it is subdominant for at least 20 years in both cases.  For the case with lower $\alpha_1$ the dust luminosities in the first few years are around $\nu L_\nu \sim 10^{37}-10^{39}$ erg/s at around $10^4-10^5$ GHz depending on the case, which would be visible within $\sim$ 100-1000 Mpc using 2500 s observations from either Spitzer or JWST.  It is also worth noting that the dust emission is not significant below $10^3$ GHz, so PWN observations with ALMA (100-250 GHz) should not be significantly affected by dust. Thus results on the detectability of non-thermal submm emission \citep{MKM16,2018MNRAS.474..573O} are unaffected. 
If $\epsilon_e < 1$, then the luminosity of the dust emission will decrease, but so will its peak wavelength, so its relative luminosity compared to the PWN spectrum will increase, possibly making the emission detectable for close supernovae even with $\alpha_1$ $\sim$ 1.8.

\subsection{Applications to Previous SNe}

SN1987A is the most well studied supernova to date, in part because of the dust formed in its ejecta.  The explosion of the $\sim$ 18-20 M$_{\sun}$ blue supergiant Sk-69 202 produced an explosion with $\sim$ 15 M$_{\sun}$ of ejecta \citep{1987Natur.328..318G}.  The ejecta contained a $\sim$ 10 M$_{\sun}$ hydrogen envelope with a $\sim$ 5 M$_{\sun}$ core of heavier elements \citep{1988PASAu...7..355W, 2006NuPhA.777..424N}.  Mg and Si were both produced in roughly equal amounts of about 0.1 M$_{\sun}$, while about 0.15-0.25 M$_{\sun}$ of C was produced \citep{1990ApJ...349..222T, 2011Sci...333.1258M}, giving mass fractions of 0.007 and 0.01-0.02 respectively; both of these are a factor of $\sim$ 5 lower than in our Ib composition.  There is evidence for the formation of both carbon and silicate dust, with almost all of the carbon gas ending up in dust grains, and about 0.4 M$_{\sun}$ of MgSiO$_3$ produced \citep{2011Sci...333.1258M, 2015ApJ...810...75D, 2015A&A...575A..95S}.  There were no silicate lines detected in the early spectra \citep{1993ApJS...88..477W}, but that could be because the emission features were absorbed by the carbon dust \citep{2015ApJ...810...75D}.  There has not yet been any detection of a compact remnant, although a pulsar with initial spin $P >$ 100 ms and $B \sim$ 10$^{11-12}$ G is still not ruled out \citep{2007AIPC..937..134M}.  Dust was hypothesized to condense in the ejecta between 415 to 615 days \citep{1993ApJS...88..477W} or even at timescales longer than 1000 days \citep{2015A&A...575A..95S, 2015MNRAS.446.2089W}, and this timescale is longer than expected for a no pulsar system; even though the C dust concentration is a factor of $\sim$ 5 lower than in our Ib composition, the Ic composition has a C concentration a factor of 3 higher but only condenses 3 days faster.  Based on our results, a pulsar with $P >$ 100 ms and $B \sim$ 10$^{11-12}$ can not explain the delay in dust formation, and a pulsar which could explain the delay would have produced detectable non-thermal radiation \citep{2016ARA&A..54...19M}, and would have heated the dust to over 1000 K, which is larger than any predicted model \citep{1993ApJS...88..477W, 2011Sci...333.1258M}.

The more recent 2012au presents an interesting case.  It is a Type Ib supernova with an ejecta mass of 5-7 M$_{\sun}$ \citep{2013ApJ...772L..17T}, making our Ib5-1 case a good approximation.  No dust emission has been reported yet, and after $\sim$ 1 year the spectrum was consistent with radioactive heating \citep{2013ApJ...770L..38M}, but [OIII] emission lines have been reported at 6.2 years, which may require another heating source.  \cite{2018ApJ...864L..36M} proposed a Crab-like pulsar on the basis of limiting the velocity of the PWN at late times, but the supernova was more luminous than a regular Type Ib supernova and had a kinetic energy of $\sim$ 10$^{52}$ erg \citep{2013ApJ...770L..38M}, similar to hypernovae, which requires a faster spinning central pulsar.  It is more likely that the central pulsar has a magnetic field of $\sim 10^{14}$ G and a period close to 1 or 2 ms; this would contribute to the luminosity and kinetic energy at early times and still provide energy to ionize the ejecta after more than 6 years, something not likely with a Crab-like pulsar.  This could possibly be tested by trying to observe dust emission, as ejecta with a Crab-like pulsar would likely have colder, larger dust and ejecta with a faster, stronger field pulsar would have smaller, hotter dust, or possibly none at all.  

Some other Galactic SNRs have been observed to have both dust and a neutron star or PWN, such as Kes 75 \citep{2008Sci...319.1802G, 2012ApJ...745...46T}, SNR G54.1+0.3 \citep{2010ApJ...710..309T, 2001A&A...370..570L}, Cas A \citep{2013ApJ...777...22E, 2010ApJ...713..356N}, and the Crab Nebula \citep{2012ApJ...753...72T}.  These SN produced between 0.01-1 M$_{\sun}$ of dust, and there has not yet been a reverse shock in any the SNR.  However, the initial spin period of these pulsars were all likely $>$ 10 ms, so the PWN likely did not have a strong effect on their dust formation.  However, since the dust found in the Crab Nebula was reported to be smaller than expected \citep{2009ASPC..414...43K, 2012ApJ...753...72T}, this may be evidence that the initial pulsar rotated with $P <$ 10 ms and suppressed grain growth, although this may be simply due to the dust mass being derived using an inaccurate distance to the Crab Nebula, as recent studies suggest the distance may be greater than previously thought \citep{2018AJ....156...58B,2018arXiv181112272F}.

\section{Discussion} \label{sec:dis}

\subsection{Dependence on $\gamma_b$} \label{sec:disgammab}

\begin{figure}
\includegraphics[width=\linewidth]{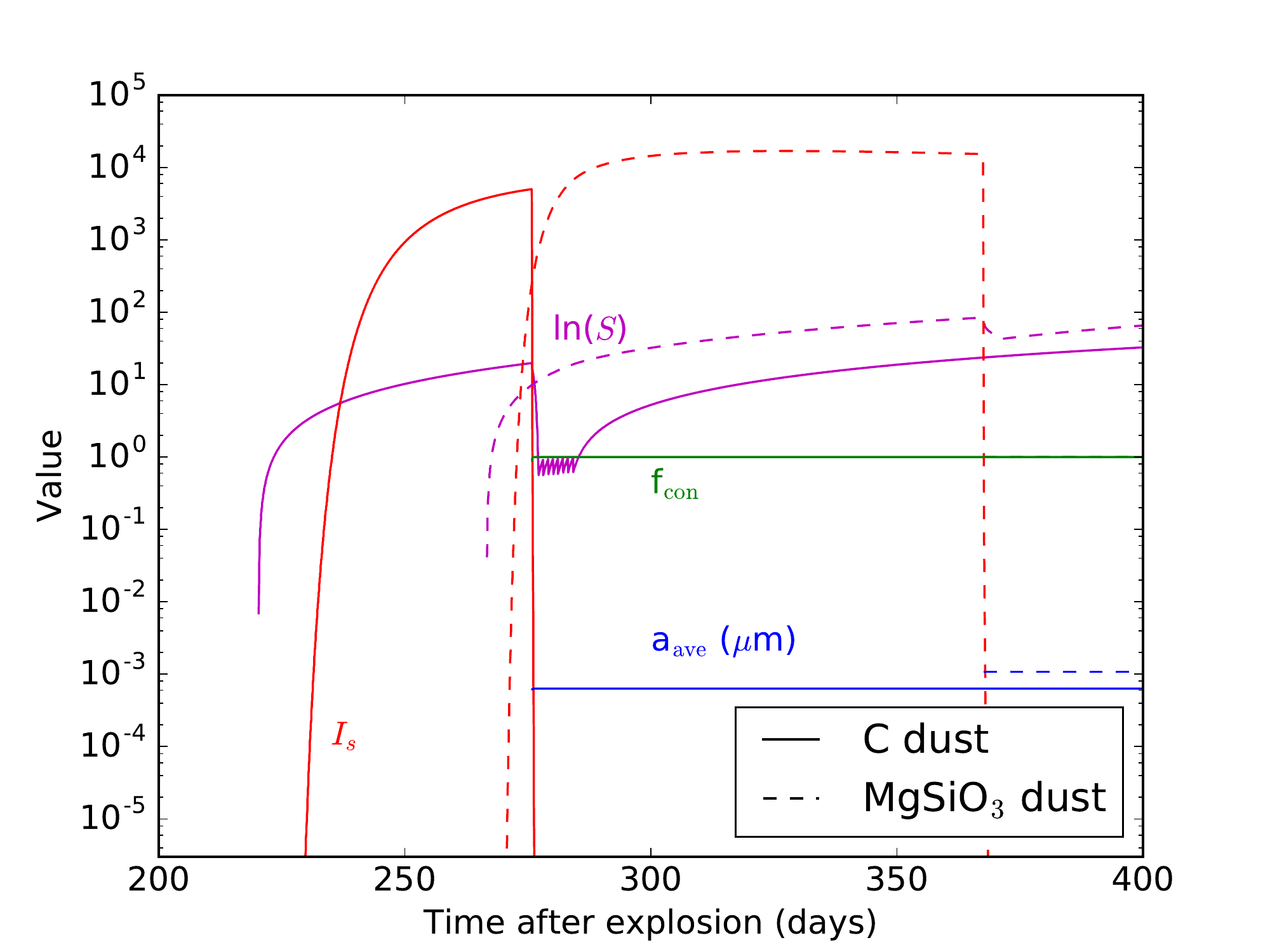}
\caption{The same as Figure~\ref{fig:ourtd} (bottom), but with $\gamma_b$ = 3 $\times$ 10$^2$ instead of 3 $\times$ 10$^5$.}
\label{fig:gbtev}
\end{figure}

The value of $\gamma_b$, which determines the break in the photon spectrum (see Equation \ref{eqn:ebsyn}), is not well constrained for very young PWNe \citep{2007whsn.conf...40V, tt13, 2014JHEAp...1...31T}.  Our value of 3 $\times$ 10$^5$ gives a spectral break in the optical to X-ray range, depending on timescale, but a value closer to 10$^2$ moves the spectral break into the submillimetre to infrared range.  We calculated the time evolution of $\ln(S)$, $I_s$, $f_{\text{con}}$ and $a_{\text{ave}}$ for the same pulsar and ejecta as Figure~\ref{fig:ourtd} (bottom), and show it in Figure \ref{fig:gbtev}.  The formation timescale is much closer to the non-pulsar case than the case with $\gamma_b$ = 3 $\times$ 10$^5$, with C dust forming around 280 days and MgSiO$_3$ forming around 370 days.  

The behaviour of the parameters for both types of dust is qualitatively similar to the pulsar cases with $\gamma_b$ = 3 $\times$ 10$^5$ cases. Both $\ln(S)$ and $I_s$ rise to a very high value before their drop at the formation time.  The drop-off in $I_s$ and rise in $f_{\text{con}}$ are also much steeper than the non-pulsar case, signifying a very short condensation timescale.  A likely interpretation of this data is that the dust was sublimated at formation at first, and the more diffuse ejecta after the temperature dropped combined with the large cluster formation rate show that as soon as the temperature dropped, all the gas immediately formed dust without having a chance to grow by further accrete gas particles.  This interpretation also explains why the average dust size is similar to the $\gamma_b$ = 3 $\times$ 10$^5$ case.

\begin{figure}
\begin{subfigure}
  \centering
  \includegraphics[width=\linewidth]{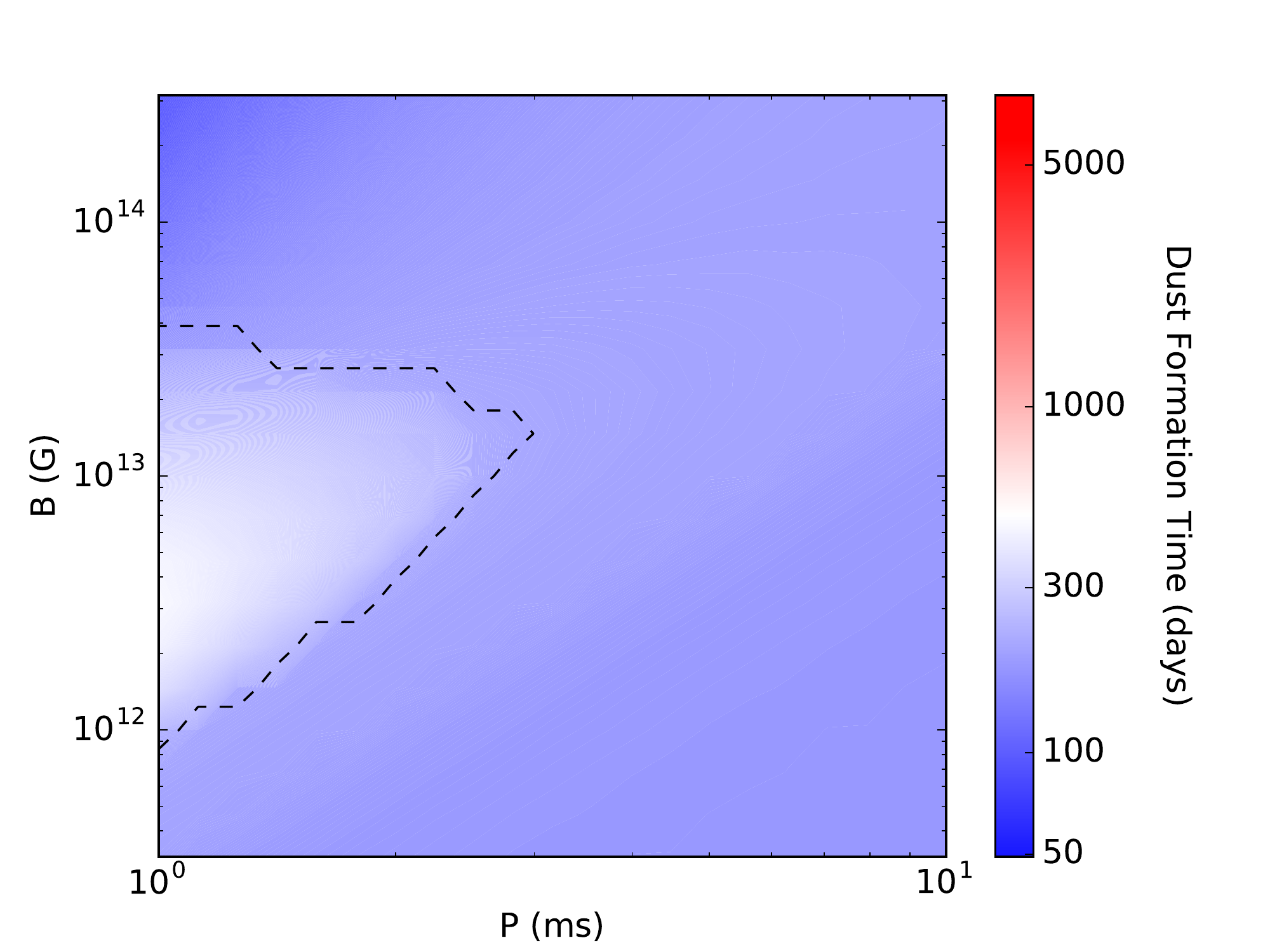}
\end{subfigure} \\
\begin{subfigure}
  \centering
  \includegraphics[width=\linewidth]{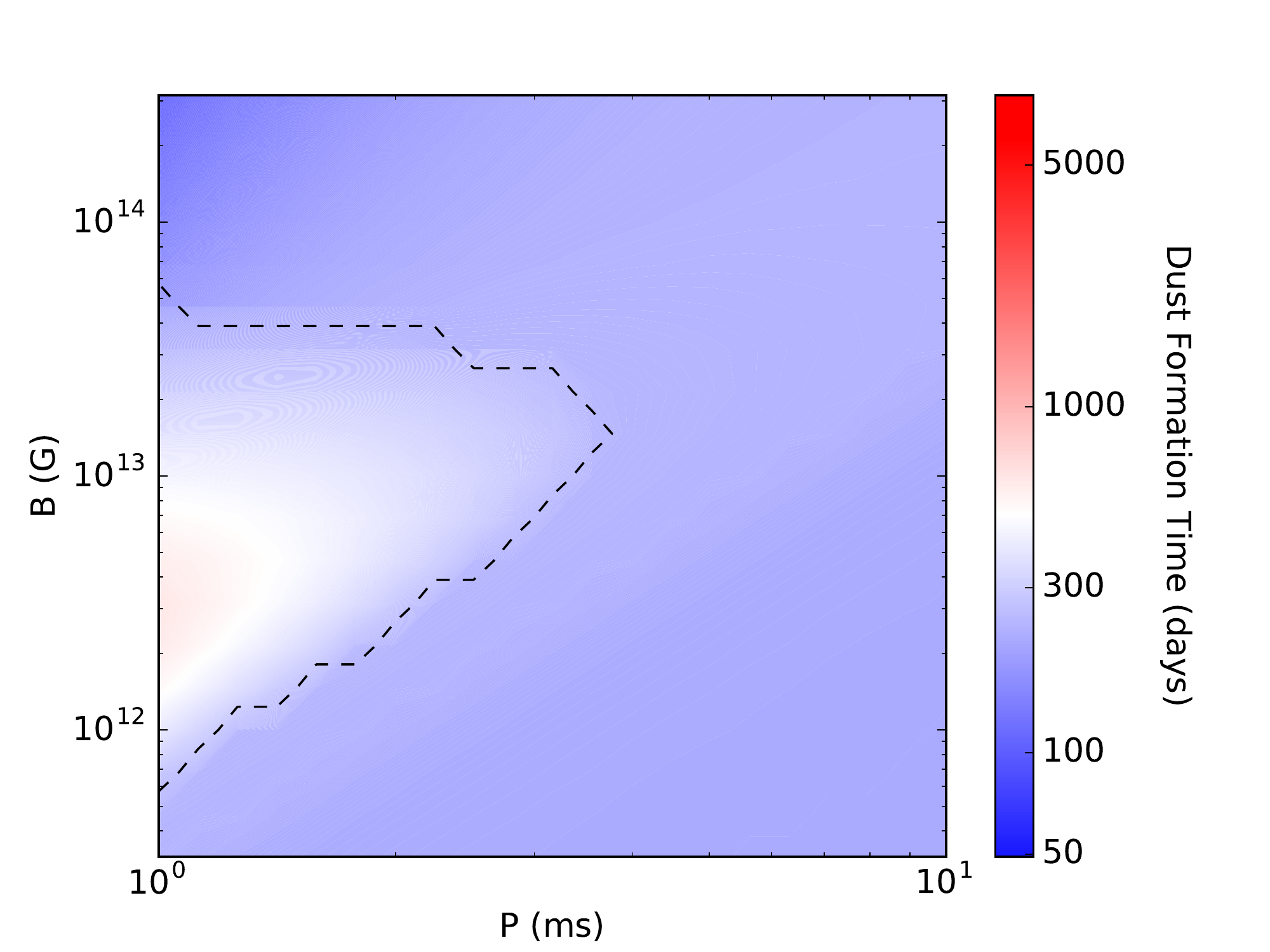}
\end{subfigure}
\caption{The formation timescale distribution for C (top) and MgSiO$_3$ (bottom) dust for the Ib5-1 parameter set with $\gamma_b$ = 3 $\times$ 10$^2$.}
\label{fig:g3e2ftd}
\end{figure}

\begin{figure}
\begin{subfigure}
  \centering
  \includegraphics[width=\linewidth]{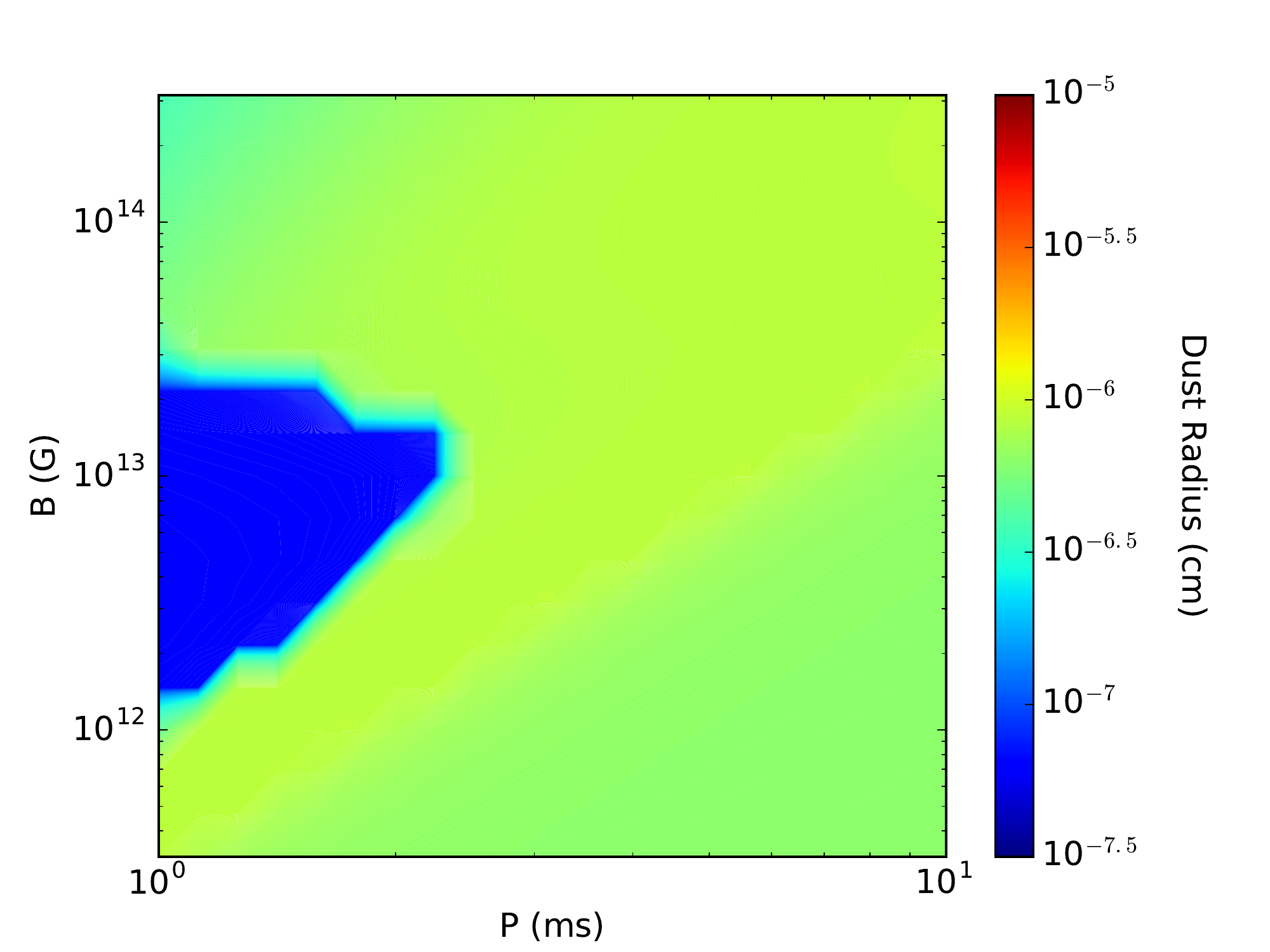}
\end{subfigure} \\
\begin{subfigure}
  \centering
  \includegraphics[width=\linewidth]{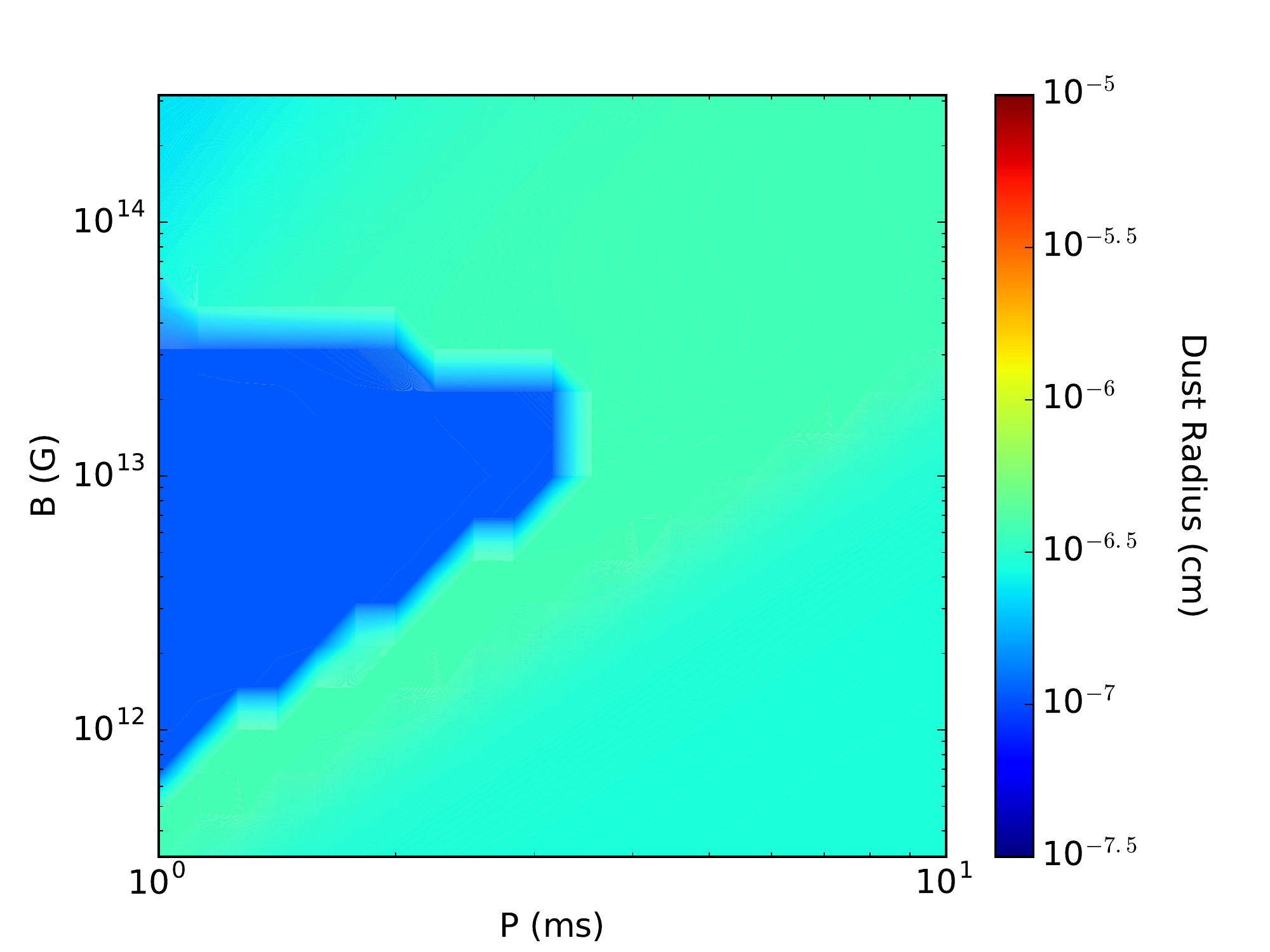}
\end{subfigure}
\caption{The average dust size distribution for C (top) and MgSiO$_3$ (bottom) dust for the Ib5-1 parameter set with $\gamma_b$ = 3 $\times$ 10$^2$.}
\label{fig:g3e2adsd}
\end{figure}

We also calculated the formation timescale distribution and average dust size distribution for the Ib5-1 parameter set, shown in Figures \ref{fig:g3e2ftd} and \ref{fig:g3e2adsd}.  We find that the effects of the pulsar are greatly reduced, with the maximum formation timescale for C and MgSiO$_3$ reduced from 1118 to 460 days and 1649 to 623 days respectively, the minimum formation timescale for C and MgSiO$_3$ increased from 58 to 93 days and 73 to 111 days respectively, ionization breakouts not occurring, and the parameter space with formation delayed by sublimation decreasing.  We also find that a 1 ms pulsar barely affects dust size for magnetar strength fields, but reduces the size around $B = 10^{12}-10^{13}$ G just as much as the $\gamma_b$ = 3 $\times$ 10$^5$ case.  

With $\gamma_b$ = 3 $\times$ 10$^2$, any trace of a pulsar engine would be difficult to detect.  The effects on formation time are much weaker than the fiducial case and occur over a smaller parameter region, and a noticeable effect on dust size is confined to a smaller region as well.  Detection of a re-emitted signal is unlikely as well, since the spectral break is now in the submillimtre/infrared, so the reprocessed emission of the dust, which will be colder due to a low optical/UV flux, will be dominated by the non-thermal emission close to the peak of the spectrum. 

\subsection{Emissivities}

We mention in Section~\ref{sec:dustsub} that the emissivities we use are the biggest uncertainties in our model, because the shape of the emissivities changes greatly with grain size \citep[e.g.,][Figures 4 and 5]{1984ApJ...285...89D}.  For C dust, the emissivity for $a > 10^{-5}$ cm is $\sim$ 1 between 1-12 eV, and decreases in this range as dust size decreases; however, the emissivity rises at higher energies at dust size decreases.  For silicates, the emissivity starts to drop off at $a >10^{-4}$~cm in the optical/UV band, but is otherwise similar to that of C.  Both types of grains absorb almost all radiation across roughly one order of magnitude, and since our spectrum is fairly flat in $\nu F_\nu$, the total energy absorbed should be comparable for all dust sizes except where $a < 10$~nm, where the emissivity drops off over the entire spectrum by a factor of 2-3.  Our bandwidth is only about half an order of magnitude so it is likely that we slightly underestimate the temperature of the dust.  The total energy absorbed should be about twice as high as our model for $a >10$~nm, so the temperature will be higher by around 1.2; this will cause sublimation to be more effective, but also make the emission more detectable.  The differences in formation time and dust size will be similar to the differences between our Ib5-05 and Ib5-1 parameter sets.  However, since we found that final grain sizes in many cases was $<$ 10 nm, this will decrease the energy absorbed by small grains by a factor of 2-3 or more, leading to sublimation being less effective for newly formed grains.

\subsection{Other Uncertainties}
Our model also has a few other caveats.  We assumes a steady state, where the current density $J_n$ from $(n-1)$-mer to $n$-mer is independent of $n$, being identical to the steady-state nucleation rate $J_s$ from Equation~\ref{eqn:jsdef}.  \cite{2013ApJ...776...24N} show that this model is only applicable if the saturation timescale $\tau_{\rm sat} \gtrsim 30\tau_{\rm coll}$, the collisional timescale, which they show to occur at higher densities; otherwise, the steady-state model condenses slightly quicker, but with smaller grains than the non-steady state.  

We assume spherical symmetry, which ignores the non-sphericity of the PWN emission, as well as the formation of clumps or fluid instabilities in the ejecta \citep{1991A&A...249..474K,2015ApJ...810...75D}.  Our calculation is based on one-zone modeling, so the dust formation rate is independent of radius; in reality, it should be sharply affected by the density profile and shell structure of the ejecta.  
Since we are interested in emission at early times, we assume no reverse shock has propagated through the ejecta; if the supernova is surrounded by circumstellar medium (CSM), the ejecta-CSM collision could send a reverse shock through the ejecta and destroy the dust \citep{2016ARA&A..54...19M, 2018MNRAS.478..110S}.

We also consider only spherical dust grains, instead of ellipsoidal grains \citep{2015ApJ...810...75D, 2015ApJ...800...50M} or more irregular shapes \citep{2003asdu.confE.170M}.  We calculate the extent of ionization in the ejecta, but we ignore the effects of the increased electron temperature and charge separation on the rest of the ejecta.  We ignore sputtering, which should decrease the average grain size \citep{1979ApJ...231...77D, 1979ApJ...231..438D, 1996ApJ...469..740J}, although the effect should not be very significant due to the thermal velocity of the grains not being extremely large \citep{1946BAN....10..187O}.  We neglect the shape of the grain distribution altogether, calculating only the average grain size; this will affect the dust emission, since even though the emission region is optically thick, there should be a range of temperatures emitted (due to different emissivities), not a single one as we assume.  

\section{Summary}

Using a model of dust formation, sublimation, emission, and gas ionization, we calculated the dust abundances, sizes, and radiation for several pulsar-driven supernovae.  We found that dust formation is qualitatively similar with and without pulsars, but it can be accelerated with $\sim$ ms rotating pulsars with super-critical magnetic fields due to the increased effectiveness of adiabatic cooling.  It can also be delayed for lower fields and higher periods due to thermal energy injection and sublimation or stopped altogether due to ionization breakout.  Carbon dust forms before silicates, and MgSiO$_3$ forms in much shorter timescales than MgO when the pulsar can delay the formation, even though they form at similar timescales when pulsars accelerate formation or do not have a significant effect on the ejecta.  Increasing ejecta mass, lowering the PWN luminosity, and lowering the key molecule concentration all delay the dust formation as well.  The typical formation timescales range from $\sim$ a few months for accelerated formation, to $\sim$ a year for no significant PWN efffect, to $\sim$ 4-6 years for delayed formation of C or MgSiO$_3$ dust, to $\sim$ 15 years for delayed formation of MgO dust.

We found that the average size of the dust is decreased to $\sim$ a few nanometers or less when a pulsar either accelerates or delays formation from the $\gtrsim$ 10 nm dust formed when pulsar energy injection is not significant, meaning that dust from pulsar-driven supernovae will likely not survive the SN reverse shock.  However, the emission from the pulsar-heated dust could be detectable out to $\sim$ 100-1000 Mpc from typical SLSNe depending on the the low-energy PWN spectral index. 

Applying this model to SNR with both dust and PWNe is not particularly insightful, as the newborn pulsar in each cases was expected to have a long enough period where PWN energy injection would not have significantly affected dust formation, although the small dust size found in the Crab Nebula could be evidence for a pulsar effect.  Applying this model to SN1987A could explain the delayed formation of dust in some models, but predicts a pulsar with non-thermal luminosity well above previous detection limits.  Applying this model to SN2012au could provide insight into the nature of the central engine depending on the possible detection and properties of dust. 

We caution that the uncertainties in the PWN spectrum at early times can greatly affect the strength of the pulsar effects.  If the value of the spectral break is greatly decreased, the effects of the pulsar are almost negligible outside of a small parameter space, where the timescale effects are weakened but the affect on dust size may be weakened or amplified. 
Our model relies on several assumptions.  The most significant assumption we make is to fix the dust absorption emissivity and assume an emission emissivity which may not be valid over the entire range of dust sizes we examine, but due to the shape of our PWN spectrum we do not expect the dust temperature to change by more than a factor of 1.2 in most cases.  Our model is also spherically symmetric, which ignores clumping and inhomogeneity in the ejecta, is one-zone, which ignores the radial dependence of concentration on dust formation, and does not fully account for effects like ionization and sputtering.  Despite this, our calculations give some insight into what emission may be expected and detectable for a pulsar-driven supernova, the timescales for which these observations may be feasible, and the fate of the dust as the SNR evolves.

\section*{Acknowledgements}
We thank Keiichi Maeda, Takaya Nozawa, and Akihiro Suzuki for discussion.  
C. M. B. O. has been supported by the Grant-in-aid for the Japan Society for the Promotion of Science (18J21778). K. K. acknowledges financial support from JSPS KAKENHI grant 18H04573 and 17K14248. K. M. acknowledges financial support from the Alfred P. Sloan Foundation and NSF grant PHY-1620777. 

\bibliographystyle{mnras}
\bibliography{ref}

\appendix

\section{Relic Electrons and Uncertainty in the PWN Spectrum} \label{sec:disrealspec}

Although the PWN spectrum we use has only one break, a more realistic one consists of multiple breaks, depending on the timescale and pulsar parameters. Detailed spectra have been studied in \cite{2015ApJ...805...82M,MKM16} taking into account electromagnetic cascades.  
In this section, we briefly review properties of non-thermal spectra. We use the notation $Q = 10^xQ_x$.

\begin{figure}
\includegraphics[width=\linewidth]{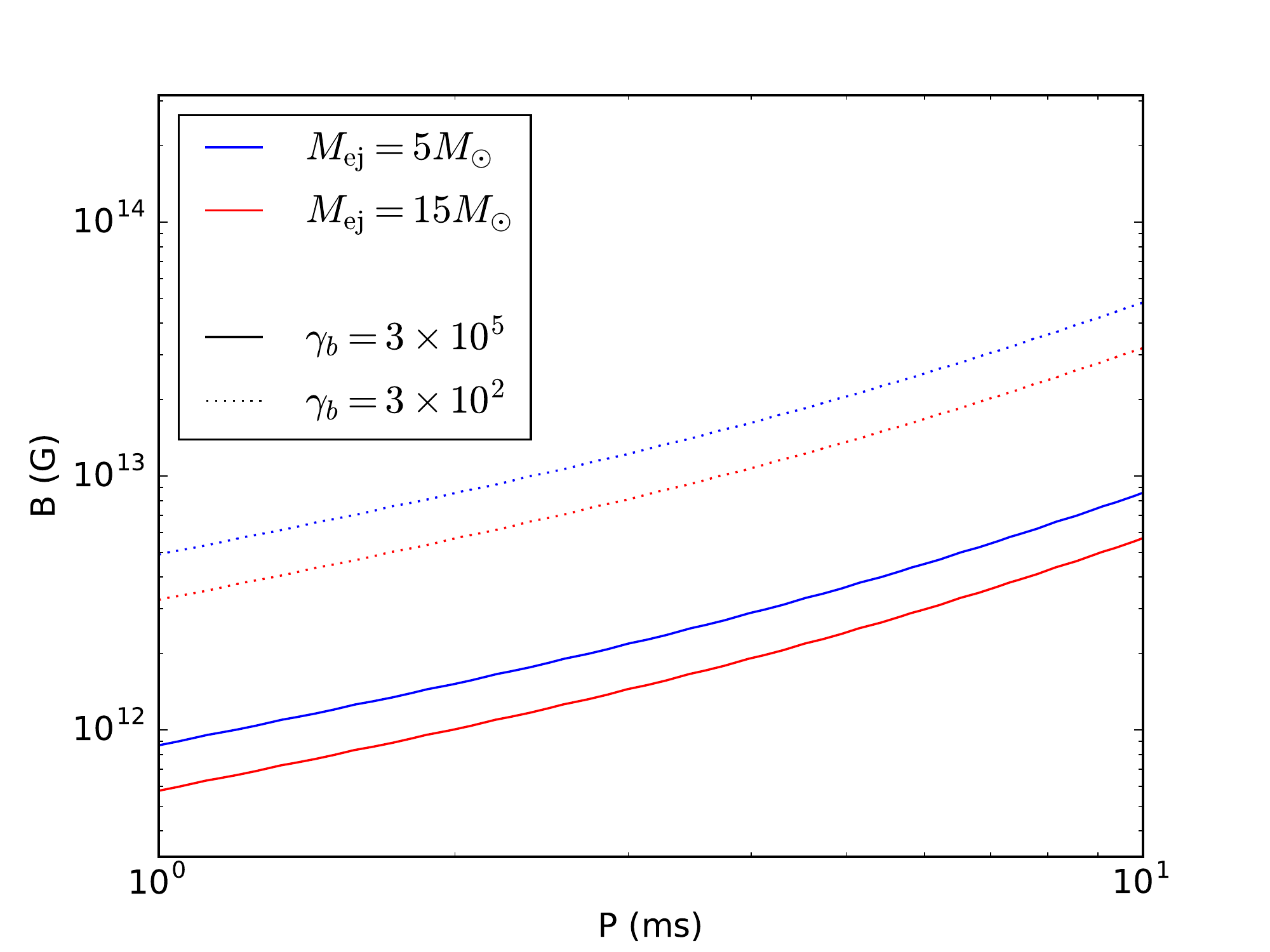}
\caption{The parameters where $t_{\rm tr} = t_{\rm SD}$.  We show the cases for both $\gamma_b$ = 3 $\times$ 10$^5$ (solid) and $\gamma_b$ = 3 $\times$ 10$^2$ (dashed) as well as for $M_{\rm ej} = 5 M_{\odot}$ (blue) and $M_{\rm ej} = 15 M_{\odot}$ (red).  $t_{\rm tr} < t_{\rm SD}$ below the line and $t_{\rm tr} > t_{\rm SD}$ above it.}
\label{fig:tsdcool}
\end{figure}

The synchrotron cooling time is
\begin{equation}
t_{\rm syn} = \frac{3 m_ec}{4\sigma_T U_B \gamma_e},
\label{eqn:tsyn}
\end{equation}
where $\sigma_T$ is the Thomson cross section, $\gamma_e$ is the electron Lorentz factor, and $U_B$ is the magnetic energy density of the PWN. 
All electrons above the cooling Lorentz factor, 
\begin{equation}
\gamma_{e,c} = \frac{\pi m_ecv_{\rm w}^3 t_{\rm SD}}{\sigma_T \epsilon_B L_{\rm SD,0}}
\begin{cases}
(t/t_{\rm SD}) & (t < t_{\rm SD}), \\
(t/t_{\rm SD})^2 & (t > t_{\rm SD}),
\end{cases}
\label{eqn:gammac}
\end{equation}
will be significantly cooled.  The spectrum is considered fast cooling if $\gamma_{e,c} < \gamma_b$ and slow cooling if $\gamma_{e,c} > \gamma_b$.  The transition time $t_{\rm tr}$ from fast to slow cooling, when $\gamma_{e,c} = \gamma_b$, is at
\begin{align}
t_{\rm tr} &= t_{\rm SD}(\gamma_b/\gamma_{e,c,\rm SD}) & \text{ if } (\gamma_{e,c,\rm SD} > \gamma_b), \\
t_{\rm tr} &= t_{\rm SD}(\gamma_b/\gamma_{e,c,\rm SD})^{1/2} & \text{ if } (\gamma_{e,c,\rm SD} < \gamma_b),
\label{eqn:ttr}
\end{align}
where the cooling Lorentz factor at the spin-down time is 
\begin{align}
\gamma_{e,c,\rm SD} &= \frac{\pi m_ecv_{\rm w}^3 t_{\rm SD}}{\sigma_T \epsilon_B L_{\rm SD,0}}, \\
&\approx 0.05 \text{ } \epsilon_{B,-3}^{-1}v_{{\rm ej},9}^{2}B_{13}^{-4}P_{-3}^{6}
\label{eqn:gammaecsd}
\end{align}
In other words, if $\gamma_{e,c,\rm SD} > \gamma_b$, then the fast to slow cooling transition happens before the pulsar spins down, and if $\gamma_{e,c,\rm SD} < \gamma_b$, the transition happens after.  Figure \ref{fig:tsdcool} shows where $\gamma_{e,c,\rm SD} = \gamma_b$ (and thus $t_{\rm tr} = t_{\rm SD}$), with the transition happening before the pulsar spins down at lower $B$ and higher $P$ and happening after spin-down at higher $B$ and lower $P$.  We show the cases for both $\gamma_b$ = 3 $\times$ 10$^5$ and $\gamma_b$ = 3 $\times$ 10$^2$ as well as for $M_{\rm ej} = 5 M_{\odot}$ and $M_{\rm ej} = 15 M_{\odot}$, as the ejecta mass affects the ejecta velocity.  Whether the pulsar parameters are above or below this line is crucial for determining what kind of PWN spectrum we expect.

Following \cite{tt10, tt13,2015ApJ...805...82M,MKM16}, we adopt broken power-law spectra for injected electrons and positrons. 
We consider spectral indices of $q_1 \sim 1-2$ and $q_2 \sim 2-3$. 
For $t < t_{\rm SD}$, the electron spectrum is dominated by freshly injected electrons, so previously cooled relic electrons are not important, but for $t > t_{\rm SD}$, adiabatically cooled relic electrons can cause a change in the spectral index~\citep{MKM16}.  

We can divide PWN spectra into 5 cases, each having different spectral breaks and indices, depending on the values of $t$, $t_{\rm SD}$, and $t_{\rm tr}$. Note that this discussion does not include low-energy synchrotron spectral breaks, such as those due to free-free absorption and synchrotron self-absorption. See \cite{MKM16} for details of the absorption effects. 

\begin{enumerate}
\item $t <  t_{\rm SD}$ and $t <  t_{\rm tr}$ - The electrons are in the fast cooling regime, and the pulsar is still spinning down: 
the synchrotron spectrum will have breaks at $\gamma_c(t)$ and $\gamma_b$ with indices of $2-\alpha= (3-q_1)/2$, $(2-q_1)/2$, and $(2-q_2)/2$ from low to high energies.

\item $t_{\rm tr} < t < t_{\rm SD}$ - The electrons are in the slow cooling regime, and the pulsar is still spinning down: 
the synchrotron spectrum will have breaks at $\gamma_b$ and $\gamma_c(t)$ with indices of $2-\alpha = (3-q_1)/2$, $(3-q_2)/2$, and $(2-q_2)/2$ from low to high energies.

\item $t_{\rm tr} < t_{\rm SD} < t$ - The pulsar has spun down and the electrons were in the slow cooling regime at $t_{\rm SD}$: 
the electron spectrum will have a break at $\gamma_b(t/t_{\rm SD})^{-1}$ due to adiabatically cooled electrons injected when the pulsar was still spinning down,   
so that the synchrotron spectrum will have breaks at $\gamma_b(t/t_{\rm SD})^{-1}$, $\gamma_b$, and $\gamma_c(t)$ with indices of $2-\alpha = (3-q_1)/2$, $1/2$, $(3-q_2)/2$, and $(2-q_2)/2$ from low to high energies.

\item $t_{\rm SD} < t < t_{\rm tr}$ - The pulsar has spun down and the electrons are still in the fast cooling regime: the electron spectrum will have a break at $\gamma_{e,c,\rm SD}(t/t_{\rm SD})^{-1}$ due to adiabatically cooled electrons injected when the pulsar was still spinning down,
so the synchrotron spectrum will have breaks at $\gamma_{e,c,\rm SD}(t/t_{\rm SD})^{-1}$, $\gamma_c(t)$, and $\gamma_b$ with indices of $2-\alpha = (3-q_1)/2$, $(7-2q_1)/3$, $(2-q_1)/2$, and $(2-q_2)/2$ from low to high energies.

\item$t_{\rm SD} < t_{\rm tr} < t$ - The pulsar has spun down and the electrons were in the fast cooling regime at $t_{\rm SD}$, but now enter the slow cooling regime:  
the synchotron spectrum will have breaks at $\gamma_{e,c,\rm SD}(t/t_{\rm SD})^{-1}$, $\gamma_b(t/t_{\rm SD})^{-1}$, $\gamma_b$, and $\gamma_c(t)$ with indices of $2-\alpha = (3-q_1)/2$, $(7-2q_1)/3$, $1/2$, $(3-q_2)/2$, and $(2-q_2)/2$ from low to high energies.
\end{enumerate}

Given the estimated values of $q_1$ and $q_2$, we find that the spectrum should only decrease in $\nu F_\nu$ above $\gamma_c(t)$ and $\gamma_b$, not just $\gamma_b$ as in our toy spectrum, with $\alpha_2$ between $2.25$ and $2.5$, which is softer than our previous $2.15$.  At lower energies, $2-\alpha$ decreases or stays constant across each break, generally taking values between 1 and 1/2, only decreasing below 1/2 between $\gamma_b$ and $\gamma_c(t)$.  

\label{lastpage}

\end{document}